\documentclass[12pt]{article}

\usepackage{epsfig}
\usepackage{psfrag}
\usepackage{latexsym}
\usepackage[DIV13]{typearea}
\usepackage{amsmath}
\usepackage{amssymb}
\usepackage{amsfonts}
\usepackage{cite}
\usepackage{bbold}
\usepackage[footnotesize]{caption2}
\usepackage{graphicx}
\usepackage[center,footnotesize,hang]{subfigure}
\usepackage{url}
\usepackage{color}

\newcommand{\mcal}[1]{\mathcal{#1}}

\newcommand{\al}{\alpha}
\newcommand{\be}{\beta}

\newcommand{\de}{\delta}
\newcommand{\De}{\Delta}

\newcommand{\La}{\Lambda}
\newcommand{\om}{\omega}
\newcommand{\sig}{\sigma}
\newcommand{\vphi}{\varphi}

\def\cC{{\cal C}}

\newcommand{\meV}{\;\text{meV}}
\newcommand{\eV}{\;\text{eV}}

\newcommand{\diag}{\text{diag}}

\newcommand{\unity}{1\hspace{-0.15cm}1}
\newcommand{\nn}{\nonumber}
\newcommand{\mean}[1]{\langle#1\rangle}

\newcommand{\beq}{\begin{equation}}
\newcommand{\eeq}{\end{equation}}
\newcommand{\bac}{\beq\begin{array}}
\newcommand{\eac}{\end{array}\eeq}
\newcommand{\ba}{\begin{array}}
\newcommand{\ea}{\end{array}}
\newcommand{\bea}{\begin{eqnarray}}
\newcommand{\eea}{\end{eqnarray}}

\newcommand{\appendixA}{\setcounter{equation}{0}
\def\theequation{\rm{A}.\arabic{equation}}\section*}
\newcommand{\appendixB}{\setcounter{equation}{0}
\def\theequation{\rm{B}.\arabic{equation}}\section*}
\begin{document}
\begin{titlepage}
\vspace*{-1cm}
\phantom{hep-ph/***}

\hfill{DFPD-09/TH/02}

\hfill{IFIC/09-04}

\vskip 1.5cm
\begin{center}
{\Large\bf Fermion Masses and Mixings in a $S_4$ Based Model}
\vskip .3cm
\end{center}
\vskip 0.5  cm
\begin{center}
{\large Federica Bazzocchi}~$^{a)}$\footnote{e-mail address: fbazzoc@few.vu.nl},
{\large Luca Merlo}~$^{b)}$\footnote{e-mail address: merlo@pd.infn.it}
\\
\vskip .2cm
and {\large Stefano Morisi}~$^{c)}$\footnote{e-mail address: morisi@ific.uv.es}
\\
\vskip .2cm
$^{a)}$~Department of Physics and Astronomy, Vrije Universiteit Amsterdam,\\
1081 HV Amsterdam, The Netherlands
\\
\vskip .1cm
$^{b)}$~Dipartimento di Fisica `G.~Galilei', Universit\`a di Padova
\\
INFN, Sezione di Padova, Via Marzolo~8, I-35131 Padua, Italy
\\
\vskip .1cm
$^{c)}$~AHEP Group, Institut de F\'{\i}sica Corpuscular --
  C.S.I.C./Universitat de Val{\`e}ncia \\
  Edificio Institutos de Paterna, Apt 22085, E--46071 Valencia, Spain
\end{center}
\vskip 0.7cm
\begin{abstract}
\noindent
It has been recently claimed that the symmetry group $S_4$ yields to the Tri-bimaximal neutrino mixing in a ``natural'' way from the group theory point of view. Approving of this feature as an indication, we build a supersymmetric model of lepton and quark masses based on this family symmetry group. In the lepton sector, a correct mass hierarchy among the charged leptons is achieved together to a neutrino mass matrix which can be diagonalized by the Tri-bimaximal pattern. Our model results to be phenomenologically unequivalent with respect to other proposals based on different flavour groups but still predicting the Tri-bimaximal mixing.
In the quark sector a realistic pattern for masses and mixing angles is obtained. The flavour structures of the mass matrices in both the sectors come from the spontaneously symmetry breaking of $S_4$, due to several scalar fields, which get non-zero vacuum expectation values. A specific vacuum alignment is required and it is shown to be a natural results of the minimization of the scalar potential and, moreover, to be stable under the corrections from the higher order terms.
\end{abstract}
\end{titlepage}
\setcounter{footnote}{0}
\vskip2truecm

%
%%%%%%%%%%%%%%%%%%%%%%%%%%%%%%%%%%%%%%%%%%%%%%%%% Introduction %%%%%%%%%%%%%%%%%%%%%%%%%%%%%%%%%%%%%%%%%%%%%%%%%%%%%%%%%%%%%%%%%%
%

\section{Introduction}

By now there is convincing evidence that the solar and the atmospheric neutrino anomalies can be explained by the neutrino oscillations. The $\Delta m^2$ values and mixing angles are known with good accuracy \cite{Data,Fogli:Indication,Maltoni:Indication}. The latest best values for $\Delta m^2$ are $\Delta m^2_{atm}\sim 2.4\times10^{-3}$ $\eV^2$ and $\Delta m^2_{sol}\sim 7.7\times10^{-5}$ $\eV^2$. For the mixing angles, two are large and one is extremely small: the atmospheric angle $\theta_{23}$ is compatible with a maximal value, but the accuracy admits relatively large deviations, indeed at $2\sigma$ errors it is $0.366 \leq \sin^2{\theta_{23}}\leq 0.602$ with central value  $0.466$; the solar angle $\theta_{12}$ is large, $0.278\leq\sin^2{\theta_{12}}\leq0.352$ with central value $0.312$, but about $5\sigma$ errors far from the maximal value; the reactor angle $\theta_{13}$ is strongly bounded and at present it has an upper limit of $\sin^2{\theta_{13}}\leq 0.036$. We underline that there are contrasting indications for a vanishing value of the reactor angle: in \cite{Fogli:Indication} there is a suggestion for a positive value which, at $1.6\sig$, is $\sin^2\theta_{13}\simeq0.016\pm0.010$, while in \cite{Maltoni:Indication} the authors find a best fit value consistent with zero within less than $1\sig$. Therefore we need a confirmation by the future experiments like DOUBLE CHOOZ \cite{Ardellier:2006mn},  Daya Bay \cite{Wang:2006ca} and MINOS \cite{PereiraeSousa:2005rf} in the $\nu_e$ appearance channel.

From the theoretical point of view, the developments about neutrino masses and mixing angles cannot satisfy: there is a so large number of existing models, that can be interpreted as a lack of a unique and compelling theoretical picture. However a series of models based on some discrete non-Abelian groups seems to be extremely attractive due to their predictions: indeed it is possible to achieve as the lepton mixing matrix the Tri-bimaximal (TB) pattern \cite{TB},
\beq
U_{TB}=\left(
         \begin{array}{ccc}
           \sqrt{\frac{2}{3}} & \frac{1}{\sqrt{3}} & 0 \\[0.3cm]
           -\frac{1}{\sqrt{6}} & \frac{1}{\sqrt{3}} & -\frac{1}{\sqrt{2}} \\[0.3cm]
           -\frac{1}{\sqrt{6}} & \frac{1}{\sqrt{3}} & \frac{1}{\sqrt{2}} \\
         \end{array}
       \right)\;,
\eeq
which represents a very good approximation of the experimental data \cite{Fogli:Indication}, providing the following values for the mixing angles:
\beq
\sin^2\theta_{13}^{TB}=0\qquad\sin^2\theta_{23}^{TB}=1/2\qquad\sin^2\theta_{12}^{TB}=1/3\;.
\eeq
It is just the TB pattern which can suggest the type of symmetry that best describes the lepton mixings: it is a very well known result \cite{af:extra} that a maximal value for the atmospheric angle can be recovered only with a non exact symmetry; explaining the indication for a non vanishing, but still very small, value for $\theta_{13}$, it is necessary to provide the TB pattern at the leading order (LO), invoking corrections from the higher order terms; the solar angle is predicted to be very close, less than $2^\circ$, to the measured value and therefore the corrections has to be relatively small. As a result, a realistic lepton flavour symmetry has to be broken at a certain level, predicting at the LO the TB pattern and providing corrections at the next-to-the-leading-order (NLO) at most of about $\theta_c^2\approx2^\circ$, where $\theta_c$ stands for the Cabibbo angle, which is a convenient hierarchical parameter for both the sectors.

There is a series of models based on the symmetry group $A_4$ \cite{TBA4,af:extra,af:modular,afl,afh,bkm,linyin,hmv}, which are extremely attractive from this point of view, fulfilling all the previous requirements. $A_4$ is the group of the even permutations of four objects and has 12 elements and four irreducible representations, which are three singlets, $1$, $1'$ and $1''$, and one triplet 3. These models manage in deriving the TB mixing by assuming that the $A_4$ symmetry is realized at a very high energy scale $\La$ and that leptons transform in a non trivial way under this symmetry. Afterward the group is spontaneously broken by a set of scalar multiplets $\phi$, the flavons, whose vacuum expectation values (VEV) receive a specific alignment. It is a non trivial task to explain how to get the expected vacuum alignment in a natural way and we consider it a fundamental requirement for a competitive model. Moreover the TB mixing is corrected by the higher order terms by quantities of the order of $\mean{\phi}/\La<1$ and as a result the reactor angle is no longer vanishing and becomes proportional to $\mean{\phi}/\La$.

The common aspect of many of these projects is the structure of the neutrino mass matrix. The most general mass matrix for the neutrinos which can be diagonalized by the TB mixing is the following
\beq
m_\nu\sim\left(
            \begin{array}{ccc}
            a+2c& b-c& b-c\\
            b-c& b+2c& a-c\\
            b-c& a-c& b+2c
            \end{array}
            \right)
\label{massaNuTB}
\eeq
and it is $\mu \leftrightarrow \tau$ invariant, yielding to a maximal atmospheric mixing angle and to a vanishing $\theta_{13}$, and it satisfies the relation $m_{\nu_{11}}+m_{\nu_{13}}=m_{\nu_{22}}+m_{\nu_{23}}$, which gives the Tri-maximal solar angle \cite{Grimus:2008tt}.
For $\theta_{13}=0$ there is no CP violation from the Dirac phase, and there are only Majorana phases. If we disregard them, we can restrict our considerations to real parameters. Usually this pattern can be obtained constructing the lagrangian in such a way that the usual Weinberg operator, which we can write as $\ell\ell$ implying $\ell h_u\ell h_u$, is forbidden at the leading order, but appears only at higher orders with additional flavons. Most of the models based on the $A_4$ flavour symmetry, are characterized by $b=0$ \footnote{The pattern with $b=0$ can be obtained with $A_4$ as flavour symmetry only with a particular flavon spectrum: the singlets $1'$ and $1''$ of $A_4$ have not to couple to the term $\ell\ell$. In fact with the left-handed leptons, $\ell$, transforming as triplet of $A_4$, we can switch on the entries corresponding to $b$ by coupling $\ell\ell$ to the flavons $F_{1'}$ and $F_{1''}$, singlets $1'$ and $1''$ respectively, as can be checked by looking in \cite{af:modular} and already underlined in the Appendix of \cite{bfm}. Moreover, in this case, the two couplings give two distinct contributions and as a result the mass matrix is not diagonalizable by the TB pattern any more. To be diagonalized by this mixing scheme, it is necessary to impose the condition $y_1\langle F_{1'}\rangle=y_2\langle F_{1''}\rangle$, where $y_i$ are the coupling constants of the two operators. However, keeping the model as natural as possible, i.e. without fine-tuning, it is necessary to prevent the couplings of the two singlets $F_{1'}$ and $F_{1''}$ with $\ell\ell$. The commonly used solution consists in not introducing such flavons.}; however, different realizations with other relations between $a,b$ and $c$ have been studied, see for instance \cite{hmv}. In the pattern with $b=0$, the factors $a$ in eq.(\ref{massaNuTB}) come from the term $\ell\ell F_1$ and the factors $c$ from $\ell\ell F_3$, where $F_1$ and $F_3$ are flavons transforming respectively as a singlet $1$ and as a triplet $3$ of $A_4$. Presenting the same flavour structure for the neutrino mass matrix, it is extremely difficult to distinguish one model from all the others by the use of only observables connected to the neutrino oscillations. Some improvements in this direction has been recently performed in \cite{fhlm:LFV}, where the authors develop an analysis on some lepton flavour violating processes, which can be tested in the future experiments.

Moreover the great difficulty of this kind of models is to describe correctly the quark sector. First of all the quark mixing matrix is completely different from its lepton counterpart: the first shows little angles and, in the contrary, the second presents two large angles. As a result, while the lepton mixing matrix can be fairly achieved through a discrete flavour symmetry, the quark mixings seem to be better described by some continuous symmetry, like $U(2)$ \cite{U(2)}. Indeed, according to the left and right-handed quark representation assignments, a discrete non-Abelian flavour symmetry tends to predict no mixing at all in the quark sector, $V_{CKM}=\unity$,  or  too large mixing angles. On the other hand, the results obtained by the $U(2)$-based models for the quarks suggest that the use of the doublet representation in the quark sector should help in describing quark mixing. However, this possibility is prevented in the $A_4$-based models, since there are not doublet representations. The solutions which have been proposed consist in the possibility of add several $Z_n$ symmetries \cite{bkm}, in order to suppress the unwanted terms, or in adopting a larger group, which manages in reproducing the structure of $A_4$ in the lepton sector and possesses some doublet representations useful to describe quarks, like for example the discrete group $T'$ \cite{fhlm:Tp,Tp}. In our opinion, a good candidate to be the flavour symmetry group describing leptons and quarks has to be as small as possible and has not to need of numerous additional elements in order to reproduce correct fermion masses and mixing angles. Following this prejudice, we looked for a proposal which manages in describing both the sectors in a realistic way, keeping as simple as possible the symmetry content.

It has been recently claimed \cite{Lam:S4natural}, through group theoretical arguments, that the minimal flavour symmetry naturally related to the TB mixing is $S_4$\footnote{We agree with the conclusions of the group theoretical analysis, but, in our opinion, it must not be considered a constraint for the model realization: from the model building point of view, the most economical realization which naturally provides the TB pattern as the neutrino mixing matrix is based on the $A_4$ symmetry group.}\cite{bm,SU5xS4}. The group $S_4$ is the group of the permutations of four objects and it has 24 elements divided into five irreducible representations: two singlets $1_1$ and $1_2$, one doublet 2 and two triplets $3_1$ and $3_2$ (a more detailed description of $S_4$ can be found in Appendix A). We approve of the result in \cite{Lam:S4natural} as an indication and we present a model based on the discrete non-Abelian symmetry group $S_4$, which predicts the TB mixing in the lepton sector and a CKM matrix close to the experimental one in the quark sector (the group $S_4$ has already been studied in
literature \cite{S4Old}, but with different aims and different results). Moreover we introduce an additional $Z_5$ symmetry, which plays a similar role of the total lepton number avoiding some dangerous terms, and a continuous $U(1)_{FN}$ \cite{FroggatNielsen}, that helps to provide the correct fermion hierarchies. $S_4$ contains as a subgroup $A_4$, but it has a doublet representation, which can be used in order to describe quarks. It has the same number of elements of $T'$, but the representations are different: in particular $T'$ can derive only the same neutrino mass matrix of the $A_4$-based models. On the other hand, the mass matrix which can be constructed in a $S_4$-based model is exactly that one in eq.(\ref{massaNuTB}) and therefore it is more general with respect to the previous case. From this point of view we can say that $T'$ constraints the neutrino sector in a stricter way than $S_4$. We underline that the model cannot be embedded into a GUT context, because of the different transformation properties of leptons with respect to quarks \footnote{When we were completing our work, the following paper appeared \cite{SU5xS4}, in which the authors present a model based on the symmetry group $SU(5)\times S_4$. However in this model it is not possible to explain completely the VEV alignment and as a result the mass hierarchies and some mixings have to be fine-tuned.}.

The presence of the doublet representation, not only represents the new expedient in order to describe the quark sector, but also introduces a new feature in the neutrino mass matrix: indeed the terms which contribute to $m_\nu$ are $\ell\ell F_1$, $\ell\ell F_3$ and the new $\ell\ell F_2$, where $F_2$ represents a flavon transforming as a doublet $2$. In eq.(\ref{massaNuTB}), this last contribution is represented by the term $b$. This result corresponds to the neutrino mass matrix in \cite{bm}: however the presence of three parameters in order to describe three masses prevents any predictions on the neutrino hierarchy type. For these reasons we conclude that the $S_4$-based model in which a singlet $F_1$, a doublet $F_2$ and also a triplet $F_3$ couple to $\ell\ell$ is not phenomenologically interesting. However it is not restrictive to construct a model in which only a singlet and a doublet contribute to the neutrino mass matrix, but in this case $m_1=m_3$ and it would be spoiled out by the experimental observations. Moreover it is possible to think about a model in which only a singlet and a triplet contribute to the neutrino mass matrix: we have verified that such a model can be built, with a natural vacuum alignment. This model provides exactly the neutrino mass matrix with $b=0$ and therefore it has the same predictions in the lepton sector as of the $A_4$-based models. For this reason in this paper we study the case in which only a doublet and a triplet couple to the term $\ell\ell$ and as a result we get an unusual neutrino mass matrix
\beq
m_\nu\sim\left(
            \begin{array}{ccc}
            2c& b-c& b-c\\
            b-c& b+2c& -c\\
            b-c& -c& b+2c
            \end{array}
            \right)
\label{massaNuS4}
\eeq
which can still be diagonalized by the TB mixing. This new pattern provides different predictions for the $0\nu2\be$-decay and thus this model can be distinguished from all the others which predict the TB mixing, just looking at some observables related to the neutrino oscillations.\\
\\
In the following we first provide a phenomenological analysis of the new neutrino mass matrix, underlining the connections with the $0\nu2\be$-decay. Subsequently, in section 3, we present the model which naturally develops the TB mixing in the lepton sector and an acceptable CKM matrix in the quark sector in addition to realistic mass hierarchies between all the fermions. In section 4, we show how to get in a natural way the special vacuum alignment, used throughout the paper. In section 5, we present a study on the corrections introduced by the higher order terms. Finally, we summarize the results in the conclusions. Details on the group $S_4$, like the conjugacy table, the complete list of the elements in a particular basis of the generators and the respective Clebsch-Gordan coefficients, can be found in Appendix A. The complete NLO analysis of the vacuum stability is presented in Appendix B.

%
%%%%%%%%%%%%%%%%%%%%%%%%%%%%%%%%%%%%%%%%%%%%%%%%% Phenomenological Analisys %%%%%%%%%%%%%%%%%%%%%%%%%%%%%%%%%%%%%%%%%%%%%%%%%%%%%%%%%%
%

\section{Phenomenological Analysis}

\begin{figure}[h!]
\begin{center}
\includegraphics[angle=0,width=8cm]{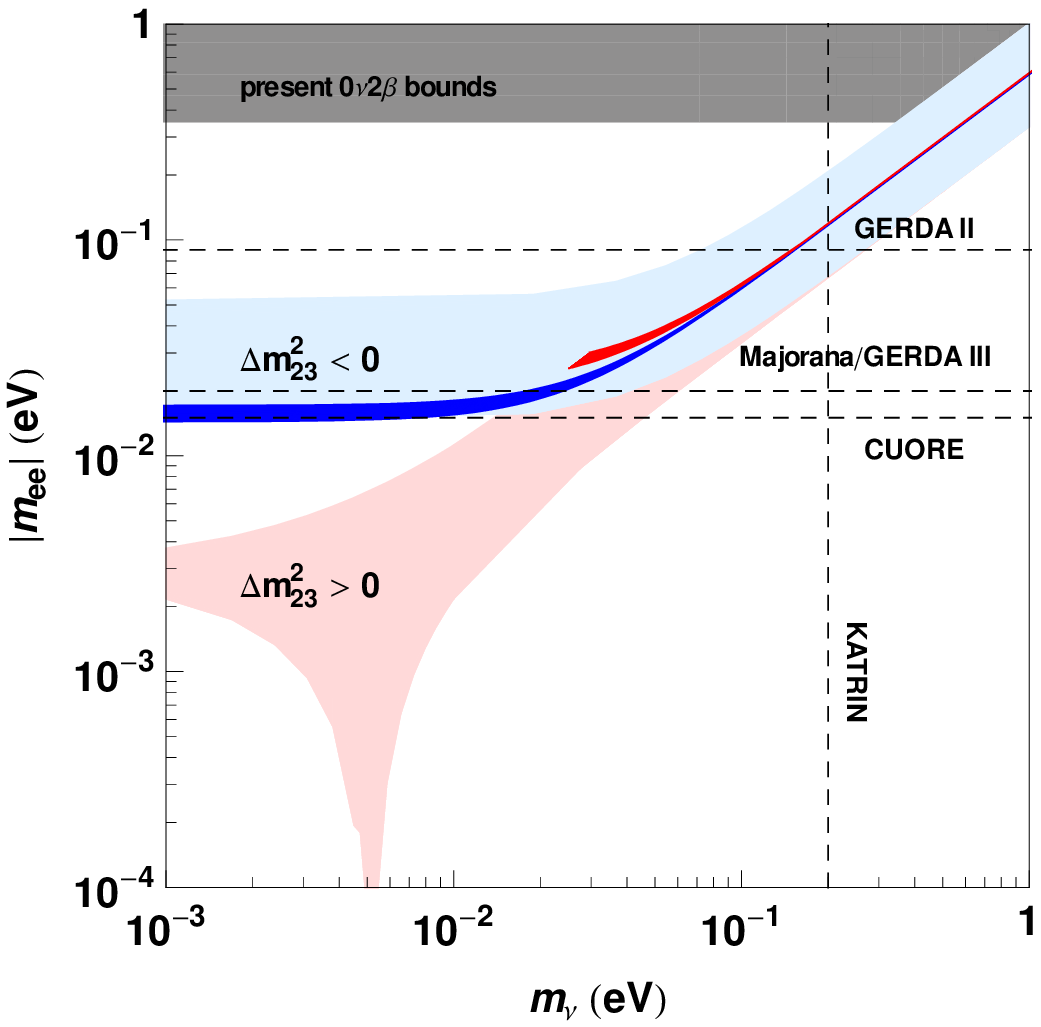}
\includegraphics[angle=0,width=8cm]{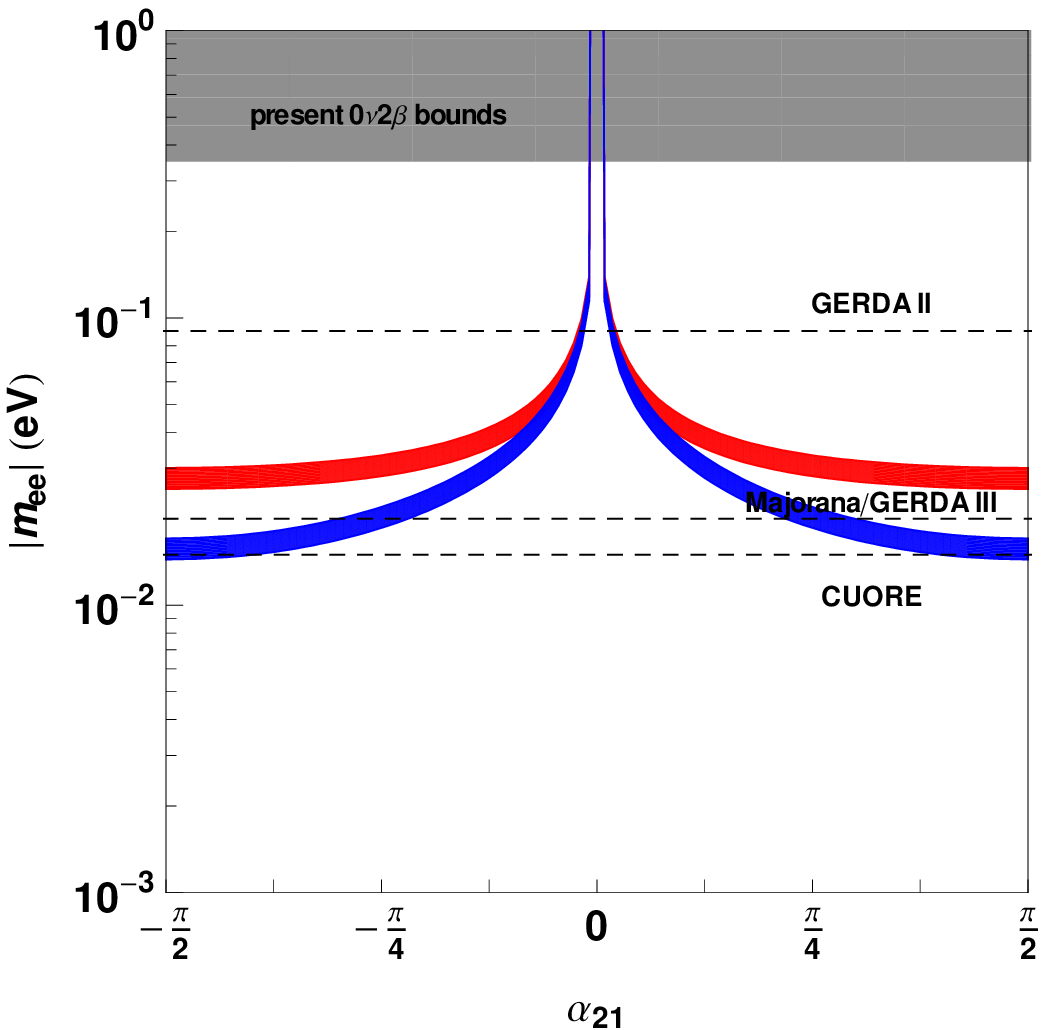}
\caption{On the left it is plotted $|m_{ee}|$ as a function of the lightest neutrino mass, $m_{\nu_1}$ in red in the case of the NH and $m_{\nu_3}$ in blue in the case of the IH. The light colored bands represent the possible regions considering only the exact TB pattern, while the dark colored ones are the predictions of our model. The present bound from the Heidelberg-Moscow experiment is shown in dark gray and the future sensitivity of CUORE ($\sim15$ meV), Majorana ($\sim20$ meV), and GERDA ($\sim90$ meV) experiments are represented by the horizontal dashed lines, while the future sensitivity of $0.2$ eV of KATRIN experiment is shown by the vertical dashed line. On the right, $|m_{ee}|$ as a function of the physical Majorana phase $\alpha_{21}$: in red the NH case and in blue the IH one. The dark gray region and the dashed horizontal lines corresponds to the previous plot.}
\label{fig1}
\end{center}
\end{figure}

The neutrino mass matrix in eq.(\ref{massaNuS4}) can be diagonalized by the TB mixing and the eigenvalues are given by
\beq
m_\nu^{diag}=(3c-b,\;2b,\;3c+b)\dfrac{v_u^2}{\La}\;.
\eeq
We can now write the neutrino oscillation parameters $\De m_{atm}^2$ and $\De m_{sol}^2$ as follows:
\begin{eqnarray}
&\Delta m_{atm}^2&=|m_{\nu_3}|^2-|m_{\nu_1}|^2=12|b||c|\cos \zeta\dfrac{v_u^4}{\La^2}\\[0.3cm]
&\Delta m_{sol}^2&=|m_{\nu_2}|^2-|m_{\nu_1}|^2=3|b|^2-9|c|^2+6|b||c|\cos \zeta\dfrac{v_u^4}{\La^2}
\end{eqnarray}
where the angle $\zeta$ is the relative phase between $b$ and $c$. This phase is related to the Majorana CP phase $\al_{21}$, which is defined as follows
\beq
U_\nu=U_{TB}\cdot\diag\left(1,e^{i\frac{\al_{21}}{2}},e^{i\frac{\al_{31}}{2}}\right)\;.
\eeq
We can express $|b|$ and $|c|$ as functions of $\Delta m_{atm}^2$, $\Delta m_{sol}^2$ and $\zeta$ and as a result we get constraints on the type of the neutrino spectrum, on the value of the lightest neutrino mass and on the $0\nu2\be$ parameter $|m_{ee}|$ directly from the experimental data. In fig.(\ref{fig1}) on the left, we plot $|m_{ee}|$ as a function of the lightest neutrino mass eigenstate, $m_{\nu1}$ in the normal hierarchy (NH) case and $m_{\nu3}$ in the inverse hierarchy (IH) one. On the right, we present $|m_{ee}|$ as a function of the Majorana phase $\al_{21}$. We observe from this last plot that considering the Heidelberg-Moscow \cite{HM} experiment, which provides the lowest present bound on $|m_{ee}|$ of about $0.35$ eV, the exact CP-conserving Majorana phase $\phi_{21}=0$ is excluded in our model. Moreover, from fig.(\ref{fig1}) on the left, we conclude that the NH region falls in the quasi Degenerate Case (DC) band and therefore we cannot speak properly of NH in this model.

Restricting our discussion to the IH case, we find a lower bound for the $0\nu2\be$ parameter, $|m_{ee}|> 14.4\;\meV$, for the lightest neutrino mass, $|m_{\nu_3}|>0.72\;\meV$, and for the sum of the neutrino masses, $\sum_i|m_i|>89.4\;\meV$, all of them corresponding to $\zeta=0$. Moreover we have a prediction for $|m_{ee}|$ in function of $\De m^2_{atm}$, $r\equiv\De m^2_{sol}/\De m^2_{atm}$ and $\zeta$
\beq
|m_{ee}|^2=\dfrac{1}{36}\left[-(1+r)\De
m^2_{atm}+\sqrt{(\De
m^2_{atm})^2\cos^2\zeta\left(3(r-1)^2+(r+1)^2\cos^2\zeta\right)}\sin^2\zeta\right]\;.\nn
\eeq
We observe that our model can be distinguished from that one in \cite{af:extra,af:modular},
based on $A_4$, and that one in \cite{fhlm:Tp}, based on $T'$, looking to the lower bound on $|m_{ee}|$:
in fact those models predict a lower bound for $|m_{ee}|$, which is about $0.005\textrm{ eV}$.
Our predictions are quite close to the future experimental sensitivity, which are expected to reach the values
of $0.090\textrm{ eV}$\cite{gerda} (GERDA), $0.020\textrm{ eV}$\cite{majorana} (Majorana), $0.050\textrm{ eV}$\cite{supernemo}
(SuperNEMO), $0.015\textrm{ eV}$\cite{cuore} (CUORE) and $0.024\textrm{ eV}$\cite{exo} (EXO).
%
%%%%%%%%%%%%%%%%%%%%%%%%%%%%%%%%%%%%%%%%%%%%%%%%% The Model %%%%%%%%%%%%%%%%%%%%%%%%%%%%%%%%%%%%%%%%%%%%%%%%%%%%%%%%%%%%%%%
%

\section{The Model}

The discrete group $S_4$ is given  by the permutations of four objects and it is composed by 24 elements. It can be defined by  two generators  $S$ and $T$ that satisfy
\begin{equation}\label{rel}
 S^4= T^3=  (ST^2)^2=\unity \,.
 \end{equation}
The  three relations reported above directly  indicate which are the discrete Abelian subgroups of $S_4$: $Z_4,Z_3,Z_2$ respectively.  Indeed the 24 elements of $S_4$ belong to five classes reported in Appendix A: the elements of $\mathcal{C}_{2,4} $ define two different sets of $Z_2$ subgroups of $S_4$, corresponding to $S^2$ and $S T^2$ respectively, those of the class $\mathcal{C}_{4}$ a set of $Z_3$ Abelian discrete symmetries associated to $T$ and those belonging to $\mathcal{C}_{5}$ a set of $Z_4$ Abelian discrete symmetries corresponding to $S$. From the three relations that define the group $S_4$ we see that it contains also a non-Abelian subgroup, $S_3$. Indeed defining $S'= S^2$ and using $S^2  T S^2= T^2$ we get the relations that define $S_3$, namely
\begin{equation}
T^3=  S^{'2}= (S' T)^2=1\,.
\end{equation}
Furthermore, $S_4$  presents $5$ irreducible representations: two singlets, $1_1,1_2$, one doublet, $2$, and two triplets, $3_1$ and $3_2$. All the technical details are reported in Appendix A.

\subsection{The Lepton Sector}

In this part we illustrate the model in the lepton sector, predicting an exact TB mixing at the LO and a realistic charged lepton mass hierarchy, by the use of flavour group $G_f$ in addition to the gauge group of the SM. The complete flavour group is $G_f=S_4\times Z_5\times U(1)_{FN}$, where the three factors play different roles: the spontaneous breaking of $S_4$ down to its subgroup $Z_2\times Z_2$ in the neutrino sector is directly responsible for the TB mixing\footnote{This breaking is extremely unusual, indeed the common preserved subgroup is $Z_2$. Here $Z_2\times Z_2$ provides the same flavour structure for the neutrino mass matrix as $Z_2$ in the $A_4$ based models and it is associated to one element of the class $\mcal{C}_2$ and one of the class $\mcal{C}_4$. The complete list of the elements are present in the Appendix A.}; the $Z_5$ factor plays a similar role of the total lepton number, avoiding some dangerous terms, and, together to the $U(1)_{FN}$, is responsible for the hierarchy among the charged fermion masses. In table \ref{table:lepton_transformation}, we can see the lepton sector fields of the model and their transformation properties under $G_f$.
\begin{table}[h]
\begin{center}
\begin{tabular}{|c||c|c|c|c|c||c||c|c||c|c||c|}
  \hline
  &&&&&&&&&&& \\[-0,3cm]
  & $\ell$ & $e^c$ & $\mu^c$ & $\tau^c$ & $h_{u,d}$ & $\theta$ & $\psi$ & $\eta$ & $\Delta$ & $\vphi$ & $\xi'$ \\
  &&&&&&&&&&& \\[-0,3cm]
  \hline
  &&&&&&&&&&& \\[-0,3cm]
  $S_4$ & $3_1$ & $1_2$ & $1_2$ & $1_1$ & $1_1$ & $1_1$ & $3_1$ & 2 & $3_1$ & $2$ & $1_2$  \\
  &&&&&&&&&&& \\[-0,3cm]
  $Z_5$ & $\om$ & $\om^3$ & 1 & $\om^2$ & 1 & 1 & $\om^2$ & $\om^2$ & $\om^3$ & $\om^3$ & 1  \\
  &&&&&&&&&&& \\[-0,3cm]
  $U(1)_{FN}$ & 0 & 1 & 0 & 0 & 0 & -1 & 0 & 0 & 0 & 0 & 0  \\
  \hline
  \end{tabular}
\end{center}
\begin{center}
\begin{minipage}[t]{12cm}
\caption{Transformation properties of the matter fields in the lepton sector and of all the flavons of the model. We distinguish the flavon fields on their role and thus we can consider $\psi$ and $\eta$ mainly connected to the charged lepton sector and $\Delta$ and $\vphi$ to the neutrino sector. All these fields together to $\xi'$ are present in the quark sector. The FN field, $\theta$, provides the correct mass hierarchy.}
\label{table:lepton_transformation}
\end{minipage}
\end{center}
\end{table}
We treat the model in a supersymmetric scenario, because the minimization of the scalar potential is simplified, but this is not a constraint from the construction of the model itself.

The superpotential for the leptons can be written as
\bea
w_\ell\;=&&\sum_{i=1}^{4}\dfrac{\theta}{\La}\dfrac{y_{e,i}}{\La^3}e^c(\ell X_i)'h_d+\dfrac{y_\mu}{\La^2}\mu^c(\ell\psi\eta)'h_d+\dfrac{y_\tau}{\La}\tau^c(\ell\psi)h_d\label{eq:wd:leptons}\\[0.3cm]
w_\nu\;=&&\dfrac{x_d}{\La^2}(\ell h_u\ell h_u\vphi)+\dfrac{x_t}{\La^2}(\ell h_u\ell h_u\Delta)\label{eq:wd:neutrinos}
\eea
where
\beq
X=\left\{\psi\psi\eta,\;\psi\eta\eta,\;\De\De\xi',\;\De\vphi\xi'\right\}
\eeq
using $(\ldots)$ to refer to the contraction in $1_1$ and $(\ldots)'$ to the contraction in $1_2$.
It is interesting to underline that the first contributions containing $e^c$ would be
\beq
\dfrac{\theta}{\La}\dfrac{y'_{e,1}}{\La^2}e^c(\ell\De\De)'h_d+\dfrac{\theta}{\La}\dfrac{y'_{e,2}}{\La^2}e^c(\ell\De\vphi)'h_d\;,
\eeq
which would dominate with respect to the terms in eq.(\ref{eq:wd:leptons}). However an explicit computation will show that these two terms are vanishing, once we assume that the flavons get this specific VEV:
\bac{rclrcl}
\mean{\psi}&=&\left(
             \begin{array}{c}
               0 \\
               1 \\
               0 \\
             \end{array}
           \right)v_\psi&
\mean{\eta}&=&\left(
             \begin{array}{c}
               0 \\
               1 \\
             \end{array}
           \right)v_\eta\\
\\
\mean{\Delta}&=&\left(
             \begin{array}{c}
               1 \\
               1 \\
               1 \\
             \end{array}
           \right)v_\Delta&
\mean{\vphi}&=&\left(
             \begin{array}{c}
               1 \\
               1 \\
             \end{array}
           \right)v_\vphi\\
\\
\mean{\xi'}&=&v_{\xi'}&
\mean{\theta}&=&v_\theta
\label{vev:allleptons}
\eac
We will demonstrate that this particular VEV alignment is a natural solution of the scalar potential in the following sections; moreover we will see that all the VEVs are of the same order of magnitude and for this reason we will parameterize the ratio $VEV/\La$ by the parameter $u$. The only VEV which originates with a different mechanism with respect to the others is $v_\theta$ and we indicate the ratio $v_\theta/\La$ by the parameter $t$.\\
With this setting, in the basis of canonical kinetic terms\footnote{It has been shown in a series of papers \cite{Kahler} that the corrections, from the transformations needed to move in the basis of canonical kinetic terms, appear at most as NLO deviations.}, the mass matrix for the charged leptons is ($m_\ell\sim R^cL$, $m_\nu\sim L^TL$)
\beq
m_\ell=\left(
         \begin{array}{ccc}
           y_e^{(1)} u^2t & y_e^{(2)} u^2t & y_e^{(3)} u^2t \\
           0 & y_\mu u & 0 \\
           0 & 0 & y_\tau \\
         \end{array}
       \right)uv_d
\eeq
where the $y_e^{(i)}$ are the result of all the different contributions of the $y_{e,i}$. For the neutrinos we get the following mass matrix, which is exactly diagonalized by the TB pattern,
\beq
m_\nu=\left(
              \begin{array}{ccc}
                2c & b-c & b-c \\
                b-c & b+2c & -c \\
                b-c & -c & b+2c \\
              \end{array}
            \right)\dfrac{v_u^2}{\Lambda}
\eeq
where $b=2x_d\dfrac{v_\vphi}{\La}$ and $c=2x_t\dfrac{v_\Delta}{\La}$. In order to find the lepton mixing matrix we need to diagonalize the charged lepton mass matrix and, performing a double expansion in the parameters $u$ and $t$, we get
\beq
m_\ell^{\diag}\equiv U_{\ell^c}^\dag m_\ell U_{\ell}=(y_e u^2t,\;y_\mu u,\;y_\tau)uv_d\;,
\eeq
where the unitary $U_\ell$ results to be the unity matrix, apart negligible corrections of order $u^2\,t^2$. As a consequence we get that
\beq
U_{PMNS}\equiv U_{\ell}^\dag U_{TB} = U_{TB}\;.
\eeq
When we introduce the NLO terms in the Lagrangian and the corrections in the VEVs of the flavons, we expect corrections to the TB mixing of relative order u, as we will discuss in the next sections. As a consequence, it provides an upper bound on the parameter $u$, indeed the maximum deviation from the TB pattern, which we can accept, is $0.05$. For $u>0.05$ the model provides a $\theta_{12}$ angle which is not in agreement at $2\sig$ error with respect to the experimental data. The hierarchy of the charged leptons comes directly from the symmetry of the model and it is possible to get a constraint on the parameters $u$ and $t$: indeed, for a very low $\tan\be$ value, the requirement for the Yukawa of the $\tau$ lepton to be in the perturbative regime ($y_\tau<4\pi$) corresponds to a lower bound for $u$ of about $0.001$. However, using this value for $u$, we require a particularly large value for $y_\mu$ in order to fulfill the measured value for the ratio $m_\mu/m_\tau$: from the requirement that also $y_\mu$ remains in the perturbative regime, the lower bound on $u$ is raised and we fix it at 0.01. In order to explain the ratio $m_e/m_\mu$, we get a range of values for the parameter $t$, which is to be similar to that for $u$. Finally we can write
\beq
0.01<u,t<0.05\,.
\label{vev:uet}
\eeq

\subsection{The Quark Sector}
In this part we illustrate the model in the quark sector, getting a good approximation of the experimental quark mixing matrix. In table \ref{table:quark_transformation}, we can see the quark sector fields of the model and their transformation properties under $S_4\times Z_5\times U(1)_{FN}$.
\begin{table}[h]
\begin{center}
\begin{tabular}{|c||c|c|c|c|c|c|c|c||c||c|c||c|c||c|}
  \hline
  &&&&&&&&&&&&&& \\[-0,3cm]
  & $D_q$ & $q_3$ & $u^c$ & $d^c$ & $c^c$ & $s^c$ & $t^c$ & $b^c$ & $\theta$ & $\psi$ & $\eta$ & $\Delta$ & $\vphi$ & $\xi'$ \\
  &&&&&&&&&&&&&& \\[-0,3cm]
  \hline
  &&&&&&&&&&&&&& \\[-0,3cm]
  $S_4$ & $2$ & $1_1$ & $1_2$ & $1_2$ & $1_2$ & $1_2$ & $1_1$ & $1_2$ & $1_1$ & $3_1$ & $2$ & $3_1$ & 2 & $1_2$  \\
  &&&&&&&&&&&&&& \\[-0,3cm]
  $Z_5$ & $\om^4$ & $\om^3$ & $1$ & $1$ & $\om^2$ & $\om^2$ & $\om^2$ & $\om^2$ & 1 & $\om^2$ & $\om^2$ & $\om^3$ & $\om^3$ & 1 \\
  &&&&&&&&&&&&&& \\[-0,3cm]
  $U(1)_{FN}$ & 0 & 0 & 2 & 1 & 0 & 0 & 0 & 0 & -1 & 0 & 0 & 0 & 0 & 0  \\
  \hline
  \end{tabular}
\end{center}
\begin{center}
\begin{minipage}[t]{12cm}
\caption[]{Transformation properties of all the fields in the quark sector.}
\label{table:quark_transformation}
\end{minipage}
\end{center}
\end{table}
The superpotential in the quark sector can be written as
\bac{rl}
w_q\;=&y_tt^cq_3h_u+\dfrac{y_b}{\La}b^cq_3\xi'h_d+\\[0.2cm]
&+\sum_{i=1}^{2}\dfrac{y_{tc,i}}{\La^2}t^c\left(D_qX^{(1)}_i\right)h_u+ \sum_{i=1}^{2}\dfrac{y_{bs,i}}{\La^2}b^c\left(D_qX^{(1)}_i\right)'h_d+\\[0.3cm]
&+\sum_{i=1}^{6}\dfrac{y_{tu,i}}{\La^3}t^c\left(D_qX^{(2)}_i\right)h_u+ \sum_{i=1}^{6}\dfrac{y_{bd,i}}{\La^3}b^c\left(D_qX^{(2)}_i\right)'h_d+\\[0.3cm]
&+\sum_{i=1}^{2}\dfrac{y_{c,i}}{\La^2}c^c\left(D_qX^{(1)}_i\right)'h_u+
\sum_{i=1}^{2}\dfrac{y_{s,i}}{\La^2}s^c\left(D_qX^{(1)}_i\right)'h_d+\\[0.3cm]
&+\dfrac{y_{ct}}{\La}c^cq_3\xi'h_u+
\sum_{i=1}^{3}\dfrac{y_{sb,i}}{\La^2}s^cq_3\left(\eta\varphi\right)'h_d+\\[0.3cm]
&+\sum_{i=1}^{6}\dfrac{y_{cu,i}}{\La^3}c^c\left(D_qX^{(2)}_i\right)'h_u+
\sum_{i=1}^{6}\dfrac{y_{sd,i}}{\La^3}s^c\left(D_qX^{(2)}_i\right)'h_d+\\[0.3cm]
&+\sum_{i=1}^{2}\dfrac{y_{u,i}}{\La^2}\dfrac{\theta^2}{\La^2}u^c\left(D_qX^{(4)}_i\right)h_u+ \sum_{i=1}^{2}\dfrac{y_{d,i}}{\La^2}\dfrac{\theta}{\La}d^c\left(D_qX^{(4)}_i\right)h_d\\[0.3cm]
&+\sum_{i=1}^{4}\dfrac{y_{ut,i}}{\La^3}\dfrac{\theta^2}{\La^2}u^c\left(q_3X^{(5)}_i\right)h_u+ \sum_{i=1}^{4}\dfrac{y_{db,i}}{\La^3}\dfrac{\theta}{\La}d^c\left(q_3X^{(5)}_i\right)h_d
\eac
where
\bea
&&X^{(1)}=\left\{\eta\eta+\psi\psi\right\}\nn\\[0.3cm]
&&X^{(2)}=\left\{\eta\eta\xi',\;\psi\psi\xi',\;\De\De\De,\;\De\De\vphi,\;\De\vphi\vphi,\;\vphi\vphi\vphi\right\}\nn\\[0.3cm]
&&X^{(3)}=\left\{\psi\De,\;\eta\vphi,\;\xi'\xi'\right\}\nn\\[0.3cm]
&&X^{(4)}=\left\{\vphi\vphi,\;\De\De\right\}\nn\\[0.3cm]
&&X^{(5)}=\left\{\psi\psi\De,\;\psi\psi\vphi,\;\psi\eta\De,\;\eta\eta\vphi\right\}\nn\;.
\eea
With this setting, the mass matrix for the up quarks is
\beq
m_u=\left(
         \begin{array}{ccc}
           y_u u^2t^2 & y_u u^2t^2 & y_{ut}u^3t^2 \\
           y_{cu} u^3 & y_cu^2 & y_{ct}u \\
           y_{tu} u^3 & y_{tc}u^2 & y_t \\
         \end{array}
       \right)v_u\;,
\eeq
and for the down quarks is
\beq
m_d=\left(
      \begin{array}{ccc}
        y_dut & y_dut & y_{db}u^2t \\
        y_{sd}u^2 & y_su & y_{sb}u \\
        y_{bd}u^2 & y_{bs}u & y_b \\
      \end{array}
    \right)uv_d\;,
\eeq
where the Yukawas are the sum of all the different terms, which appear in the superpotential.\\
These mass matrices can be diagonalized by the following transformations:
\beq
m_u^{\diag}\equiv U_{u^c}^\dag m_u U_u=(y_uu^2t^2,\;y_cu^2,\;y_t)v_u\qquad
m_d^{\diag}\equiv U_{d^c}^\dag m_d U_d=(y_dut,\;y_su,\;y_b)uv_d
\eeq
where the unitary matrices can be written in terms of order of magnitude of $u$ and $t$ as
\bac{rlrl}
U_u=&\left(
      \begin{array}{ccc}
        1 & O(u) & O(u^3) \\
        -O(u) & 1 & O(u^2) \\
        -O(u^3) & -O(u^2) & 1 \\
      \end{array}
    \right)&
U_d=&\left(
      \begin{array}{ccc}
        1 & O(u) & O(u^2) \\
        -O(u) & 1 & O(u) \\
        -O(u^2) & -O(u) & 1 \\
      \end{array}
    \right)\\[0.8cm]
U_{u^c}=&\left(
      \begin{array}{ccc}
        1 & O(t^2) & -O(ut^2) \\
        -O(t^2) & 1 & O(u) \\
        -O(ut^2) & -O(u) & 1 \\
      \end{array}
    \right)\quad&
U_{d^c}=&\left(
      \begin{array}{ccc}
        1 & O(t) & O(ut) \\
        -O(t) & 1 & O(u) \\
        -O(ut) & -O(u) & 1 \\
      \end{array}
    \right)\;.
\eac
The resulting quark mixing matrix is
\beq
V_{CKM}\equiv U_u^\dag U_d\simeq\left(
                                  \begin{array}{ccc}
                                    1 & \left(\dfrac{y_{sd}}{y_s}-\dfrac{y_{cu}}{y_c}\right)u & \left(\dfrac{y_{bd}y_{c}-y_{bs}y_{cu}}{y_by_c}\right)u^2 \\[0.3cm]
                                    -\left(\dfrac{y_{cu}}{y_c}-\dfrac{y_{sd}}{y_s}\right)u & 1 & \dfrac{y_{bs}}{y_b}u \\[0.3cm]
                                    \left(y_{bs}y_{sd}-\dfrac{y_{bd}y_{s}}{y_by_s}\right)u^2 & -\dfrac{y_{bs}}{y_b}u & 1 \\
                                  \end{array}
                                \right)\;.
\eeq

In order to fit the experimental values of the mixing angles we need to invoke a moderate fine-tuning in some parameters. The (23) entry of $V_{CKM}$ has to be of order $\theta_c^2\simeq0.05$ and therefore suggests for $u$ a value close to its upper bound. However this is not a strict constraint because this value can be well explained for the entire range of $u$ considering the Yukawas. On the other hand, the entry (12) requires an accidental enhancement of the combination $\left(\dfrac{y_{sd}}{y_s}-\dfrac{y_{cu}}{y_c}\right)$ of order $1/\theta_c\sim4$ in order to describe the correct Cabibbo angle. It is possible to explain such an enhancement considering particular values of the relative phase, $\zeta_q$, between $\dfrac{y_{sd}}{y_s}$ and $\dfrac{y_{cu}}{y_c}$, which is connected to the CP violating phase: if $\zeta_q=\pi$, then the two factors sum up and the required values are easily explained.

%
%%%%%%%%%%%%%%%%%%%%%%%%%%%%%%%%%%%%%%%%%%%%%%%%% The Vacuum Alignment %%%%%%%%%%%%%%%%%%%%%%%%%%%%%%%%%%%%%%%%%%%%%%%%%%%%%%%%%%%%%%%
%

\section{The Vacuum Alignment}
In the following we present the mechanism to get the particular VEV alignment used in the previous sections. In table \ref{table:flavon_transformation} we illustrate all the flavon fields of the model and a set of new fields, the driving fields, defined as scalar fields with vanishing VEV, which are used only to select the particular solutions of the scalar potential. In order to distinguish between the matter fields, the flavons and the driving fields we introduce an additional $U(1)_R$, under which the fields have quantum number 1, 0 and 2 respectively. The usual R-parity, useful to avoid FCNC in the supersymmetric extensions of the SM, is a discrete group of this $U(1)_R$.
\begin{table}[ht]
\begin{center}
\begin{tabular}{|c||c|c|c|c||c|c|c||c|c|}
  \hline
  &&&&&&&&& \\[-0,3cm]
  & $\Delta$ & $\vphi$ & $\Delta^0$ & $\vphi^0$ & $\psi$ & $\eta$ & $\psi^0$ & $\xi'$ & $\xi'^0$ \\
  &&&&&&&&& \\[-0,3cm]
  \hline
  &&&&&&&&& \\[-0,3cm]
  $S_4$ & $3_1$ & 2 & $3_2$ & 2 & $3_1$ & $2$ & $3_1$ & $1_2$ & $1_2$ \\
  &&&&&&&&& \\[-0,3cm]
  $Z_5$ & $\om^3$ & $\om^3$ & $\om^4$ & $\om^4$ & $\om^2$ & $\om^2$ & $\om$ & 1 & 1 \\
  \hline
  \end{tabular}
\end{center}
\begin{center}
\begin{minipage}[t]{12cm}
\caption[]{Transformation properties of the flavons and the driving fields.}
\label{table:flavon_transformation}
\end{minipage}
\end{center}
\end{table}

The driving superpotential is
\bac{rcl}
w_d&=&g_1(\De^0\De\vphi)+g_2(\vphi^0\De\De)+g_3(\vphi^0\vphi\vphi)+\\[0.3cm]
&&+f_1(\psi^0\psi\psi)+f_2(\psi^0\psi\eta)+\\[0.3cm]
&&+M_{\xi'}\xi'^0\xi'+h_1\xi'^0(\eta\vphi)'
\label{eq:wd:driving}
\eac
The equations for the minimum of the scalar potential are obtained deriving $w_d$ by the driving fields:
\begin{subequations}
\begin{flalign}
\label{eq:wd:Neutrinos1}
&g_1(\vphi_1\De_2-\vphi_2\De_3)=0\\
\label{eq:wd:Neutrinos2}
&g_1(\vphi_1\De_1-\vphi_2\De_2)=0\\
\label{eq:wd:Neutrinos3}
&g_1(\vphi_1\De_3-\vphi_2\De_1)=0
\end{flalign}
\end{subequations}
\begin{subequations}
\begin{flalign}
\label{eq:wd:Neutrinos4}
&g_2(\De_3^2+2\De_1\De_2)+g_3\vphi_1^2=0\\
\label{eq:wd:Neutrinos5}
&g_2(\De_2^2+2\De_1\De_3)+g_3\vphi_2^2=0
\end{flalign}
\end{subequations}
\begin{subequations}
\begin{flalign}
\label{eq:wd:ChargedLeptons1}
&2f_1(\psi_1^2-\psi_2\psi_3)+f_2(\eta_1\psi_2+\eta_2\psi_3)=0\\
\label{eq:wd:ChargedLeptons2}
&2f_1(\psi_2^2-\psi_1\psi_3)+f_2(\eta_1\psi_1+\eta_2\psi_2)=0\\
\label{eq:wd:ChargedLeptons3}
&2f_1(\psi_3^2-\psi_1\psi_2)+f_2(\eta_1\psi_3+\eta_2\psi_1)=0
\end{flalign}
\end{subequations}
\beq
\label{eq:wd:Quarks}
M_{\xi'}\xi'+h_1(\eta_1\vphi_2-\eta_2\vphi_1)=0
\eeq

The equations can be divided into almost separated groups. The first five equations, (\ref{eq:wd:Neutrinos1})-(\ref{eq:wd:Neutrinos5}), are satisfied by the alignment
\beq
\mean{\De}=\left(
              \begin{array}{c}
                1 \\
                1 \\
                1 \\
              \end{array}
            \right)v_\De\qquad
\mean{\vphi}=\left(
              \begin{array}{c}
                1 \\
                1 \\
              \end{array}
            \right)v_\vphi\;,
\label{vev:neutrinos}
\eeq
which is a stable solution of the scalar potential, with
\beq
v_\De^2=-\dfrac{g_3}{3g_2}v_\vphi^2\qquad\qquad v_\vphi\;\text{undetermined}\,.
\eeq
The three equations (\ref{eq:wd:ChargedLeptons1})-(\ref{eq:wd:ChargedLeptons3}), almost separated from the others, are satisfied by two different patterns: the first is
\beq \mean{\psi}=\left(
               \begin{array}{c}
                 0 \\
                 1 \\
                 0 \\
               \end{array}
             \right)v_\psi\qquad
\mean{\eta}=\left(
              \begin{array}{c}
                0 \\
                1 \\
              \end{array}
            \right)v_\eta
\label{vev:leptons}
\eeq
with
\beq
v_\psi=-\dfrac{f_2}{2f_1}v_\eta\qquad v_\eta\text{ undetermined}
\eeq
and the second is
\beq
\mean{\psi}=\left(
                   \begin{array}{c}
                     1 \\
                     1 \\
                     1 \\
                   \end{array}
                 \right)v_\psi\qquad
\mean{\eta}=\left(
                  \begin{array}{c}
                    1 \\
                    -1 \\
                  \end{array}
                \right)v_\eta
\label{vev:leptonsno}
\eeq
with $v_\eta$ and $v_\psi$ undetermined. Only the first solution provides the results presented in the previous sections and we need of some soft masses in order to discriminate it as the lowest minimum of the scalar potential. We manage in doing it, considering some $Z_5$-breaking soft terms involving $\psi$ and $\eta$, which in the most general form can be written as
\beq
m^2_\psi |\psi|^2+ m^2_\eta |\eta|^2 +\tilde{m}^2_\psi \psi \psi  + \tilde{m}^2_\eta \eta \eta\,.
\eeq
Assuming that $m^2_{\psi,\eta} <0$ the first two terms  stabilize the potential for both the vacuum configurations. On the other hand the last two terms vanish for the first vacuum configuration and get a value different from zero in the second one.  With an apposite choice of  the soft parameters, these contributions can be positive, distinguishing the two configurations of VEVs and assuring that one in eq.(\ref{vev:leptons}) as the setting with the corresponding lowest minimum.

Acting on the configurations of eq.(\ref{vev:neutrinos}) or eq.(\ref{vev:leptons}) with elements of the flavour symmetry group $S_4$, we can generate other minima of the scalar potential. These new minima are physically equivalent to those of the original sets, but it is not restrictive to analyze the model by choosing as local minimum exactly those ones in eqs.(\ref{vev:neutrinos}) and (\ref{vev:leptons}) (it is possible to show that the different scenarios are related by field redefinitions).

The last equation (\ref{eq:wd:Quarks}) connects all the sectors and fixes the VEV of $\xi'$
\beq
\mean{\xi'}=v_{\xi'}=\dfrac{h_1}{M_{\xi'}}v_\eta v_\vphi\;.
\eeq

For the flavon field $\theta$, related to the Froggatt-Nielsen symmetry, the non vanishing VEV is determined by the D-term associated with the $U(1)_{FN}$ symmetry (see \cite{afh} for more details). The D-term in the potential is given by:
\beq
V_D=\frac{1}{2}(M^2_{FI}-g_{FN}|\theta|^2+\dots)^2
\eeq
where $g_{FN}$ is the gauge coupling constant of $U(1)_{FN}$ and $M^2_{FI}$ is the Fayet-Iliopoulos term. The vanishing of $V_D$ requires
\beq
g_{FN}|\theta|^2=M^2_{FI}.
\eeq
Assuming that $M^2_{FI}/g_{FN}$ is positive, this condition fixes the VEV of $\theta$, given in eq. (\ref{vev:uet}).~

%
%%%%%%%%%%%%%%%%%%%%%%%%%%%%%%%%%%%%%%%%%%%%%%%%% NLO Corrections %%%%%%%%%%%%%%%%%%%%%%%%%%%%%%%%%%%%%%%%%%%%%%%%%%%%%%%%%%%%%%%
%

\section{NLO Corrections}
We now study the deviations to the LO results. We first present the analysis for the VEV alignment and then we move to the mass matrices.

\subsection{The VEV Alignment}
Here we only summarize the results for the vacuum alignment, while a detailed study is developed in the Appendix B.
The part of the superpotential depending on the driving fields $\De^0$, $\vphi^0$, $\psi^0$ and $\xi'^0$ is modified into
\beq
w_d=w^0_d+\delta w_d\;,
\eeq
where $w^0_d$ corresponds to eq.(\ref{eq:wd:driving}) and $\delta w_d$ is the most general quartic, $S_4$-invariant polynomial linear in the driving fields:
\beq
\delta w_d=\dfrac{1}{\La}\left(\sum_{i=1}^5x_iI_i^{\De^0}+\sum_{i=1}^{6}w_iI_i^{\vphi^0}+\sum_{i=1}^7s_iI_i^{\psi^0}+
\sum_{i=1}^2v_iI_i^{\xi'^0}\right)
\eeq
where $x_i$, $w_i$, $s_i$ and $v_i$ are coefficients and $\left\{I_i^{\De^0},\;I_i^{\vphi^0},\;I_i^{\psi^0},\;I_i^{\xi'^0}\right\}$ represents a basis of independent quartic invariants (the list of all the $I_i^{\Phi^0}$ are present in Appendix B).
The new minimum is obtained by searching for the zeros of the F terms, looking for a solution that perturbs eq.(\ref{vev:allleptons}) to first order in the $1/\La$ expansion: denoting the general flavon field with $\Phi$, we can write the new VEVs as
\beq
\mean{\Phi_i}=\mean{\Phi_i}^{(LO)}+\de\Phi_i\;.
\eeq
All the perturbations are non vanishing, a part $\de\eta_1$ and $\de\eta_2$ and one of the perturbations in the neutrino sector, which remains undetermined. On the other hand the NLO terms fixes the relation between $v_\vphi$ and $v_\eta$. We can conclude that the VEV alignment in eq.(\ref{vev:allleptons}) is stable under the NLO corrections and the deviations are of relative order $u$ with respect the LO results.

\subsection{The Mass Matrices}
In this part we present the corrections to the mass matrices due to the higher order terms in the matter superpotential and the deviations to the VEV alignment.
\begin{description}
  \item[Lepton Sector]:
the superpotential for the charged leptons can be written as
\beq
w_\ell=w_\ell^0+\de w_\ell
\eeq
where $w_\ell^0$ corresponds to eq.(\ref{eq:wd:leptons}) and $\de w_\ell$ contains all the NLO terms. We note that the LO operators related to $e^c$ completely fill in the first line of $m_\ell$ and, as a result, the corrections can be reabsorbed in the LO parameters. For this reason, we avoid to specify the NLO operators of $\de w_\ell$ related to $e^c$, reporting only those ones connected to $\mu^c$ and $\tau^c$:
denoting $\De$ and $\vphi$ with $\Phi_\nu$ and $\psi$ and $\eta$ with $\Phi_\ell$, we can write
\beq
\dfrac{\tau^c}{\La^2}(\ell\Phi_\ell\Phi_\ell\Phi_\nu+\ell\Phi_\ell\xi'\xi')\;,\qquad\qquad
\dfrac{\mu^c}{\La^3}(\ell\Phi_\nu\Phi_\nu\Phi_\nu+\ell\Phi_\ell\Phi_\ell\xi')\;.
\eeq
These corrections have to be added to those ones originated by $w_\ell^0$ considering the deviations at the NLO to the vacuum alignment. Finally the corrected charged lepton mass matrix has the following structure
\beq
m_\ell=\left(
         \begin{array}{ccc}
           O(u^2t) & O(u^2t) & O(u^2t) \\
           O(u^2) & O(u) & O(u^2) \\
           O(u) & O(u) & O(1) \\
         \end{array}
       \right)uv_d\;,
\eeq
where only the order of magnitude of the single entries are reported. As a consequence the unitary matrix $U_\ell$, which corresponds to the transformation of the charged leptons used to diagonalized $m_\ell$, is modified in the following way:
\beq
U_\ell=\left(
         \begin{array}{ccc}
           1 & T_{12}^e u & T_{13}^e u \\
           -T_{12}^e u & 1 & T_{23}^e u \\
           -T_{13}^e u & -T_{23}^e u & 1 \\
         \end{array}
       \right)\;,
\label{eq:NLO:charged leptons}
\eeq
where the parameters $T_{ij}^e$ are factors of order one.\\

A similar analysis can be performed for the neutrino superpotential
\beq
w_\nu=w_\nu^0+\de w_\nu
\eeq
where $w_\nu^0$ corresponds to eq.(\ref{eq:wd:neutrinos}) and $\de w_\nu$ contains the only NLO operator,
\beq
\dfrac{x_d'}{\La^3}(\ell h_u\ell h_u\vphi)'\xi'\;.
\eeq
In addition to this correction, we have to consider those ones from $w_\nu^0$, with the deviations at the NLO to the VEVs.
As a consequence the neutrino mass matrix is corrected by terms of relative order $u$ in every entry. Now the TB pattern has to be modified in order to diagonalize $m_\nu$ and we can write
\beq
U_\nu=U_{TB}+\de U_\nu u
\eeq
where $\de U_\nu$ can be parameterized by three angles, $T_{12}^\nu$, $T_{23}^\nu$ and $T_{13}^\nu$, in a similar way as in eq.(\ref{eq:NLO:charged leptons}).\\

Finally, summarizing all the corrections from the higher order terms, deviations to the neutrino mixing matrix of relative order $u$ with respect the LO results are generated. The corrected neutrino mixing angles are modified as follows:
\bea
&&\tan\theta_{23}=-1-2u\left(T_{23}^e+\dfrac{\sqrt2T_{13}^\nu-2T_{23}^\nu}{\sqrt3}\right)\\[0.3cm]
&&\tan\theta_{12}=\dfrac{1}{\sqrt2}-\dfrac{3u}{4}\left(\sqrt2\left(T_{12}^e+ T_{13}^e\right)-\sqrt3\left(T_{12}^\nu+T_{13}^\nu\right)\right)\\[0.3cm]
&&\tan\theta_{13}=\dfrac{u}{2\sqrt3}\left(\sqrt6\left(T_{12}^e-T_{13}^e\right)+T_{13}^\nu+2\sqrt2T_{23}^\nu-3T_{12}^\nu\right)\;.
\eea
We can conclude that the NLO corrections originate deviations to the TB mixing angles of order $u$.
  \item[Quark Sector]: the analysis for the up and down quark mass matrices is simpler than the previous case, because $m_u$ and $m_d$ do not have any vanishing entry at LO and therefore the corrections from the NLO operators of the superpotential and from the deviations to the VEVs introduce correcting factors of relative order $u$ in each entry of the mass matrices. As a result the quark mixing angles receive deviations of relative order $u$, which do not spoil the LO results.
\end{description}

%
%%%%%%%%%%%%%%%%%%%%%%%%%%%%%%%%%%%%%%%%%%%%%%%%% Conclusions %%%%%%%%%%%%%%%%%%%%%%%%%%%%%%%%%%%%%%%%%%%%%%%%%%%%%%%%%%%%%%%
%

\section{Conclusions}
The aim of a flavour model is getting the correct mixing angles and mass hierarchies of both leptons and quarks, without inducing not observed processes, like FCNC and proton decays. Moreover in the context of non-Abelian flavour discrete symmetries, we face off the further problem of keeping and preserving a different VEV alignment for the flavons and also this requirement has to be naturally fulfilled\footnote{Some attempts in which the VEV alignment problem in not present can be found in \cite{NoAlignment}.}. All these points are separated one from each other and it seems very hard to get all of them at the same time using a single flavour symmetry group. However if a model manages in doing it, it will be considered as the most promising model in order to describe nature. Trying to understand what is the best candidate, many models have been proposed based on a product of different symmetry groups. However only few of them appear interesting: we consider fundamental aspects the lack, or at least a moderate amount, of fine-tunings, the smallness of the number of elements in the complete flavour symmetry group and in the list of the new scalar fields, the flavons.

Following these points, in this paper we have presented a model for fermion masses and mixing angles based on the flavour symmetry group $S_4\times Z_5\times U(1)_{FN}$. The main aspect is the spontaneous breaking of $S_4$, which guarantees the TB pattern as the neutrino mixing matrix at LO. This feature is common to other models, like for example those based on the group $A_4$ \cite{TBA4,af:extra,af:modular,afl,afh,bkm,linyin,hmv} or those containing $A_4$, as the group $T'$ \cite{fhlm:Tp,Tp}. The choice of $S_4$ has been suggested by the recent work by C.~S.~Lam \cite{Lam:S4natural}: $S_4$ results to be the only group (with all the groups containing $S_4$) which predicts the TB mixing in a natural way, namely without \emph{ad hoc} assumptions, from the group theory point of view. This result is completely apart from all the possible realizations of a model based on $S_4$ and predicting the TB pattern: indeed, from the model building point of view, the most economic group which realizes this particular neutrino mixing matrix is $A_4$. However, there are other reasons which enforce the use of $S_4$: in the $A_4$-based models, it seems very difficult and unnatural to generate the correct mass hierarchies and mixings for quarks. $S_4$ represents a viable solution to this problem, because it contains a doublet representation more than $A_4$, which can be used in order to describe quarks.

This is not the first attempt in this direction: in \cite{fhlm:Tp} the group $T'$ has been studied with good results, getting the TB mixing for leptons and a realistic quark mixings together to correct mass hierarchies. Unfortunately, this model suffers of a fine-tuning in order to generate the up-quark mass and the $(12)$ entry of the CKM matrix: these negative aspects have been already underlined in the papers by Barbieri \emph{et al.}, studying the continuous group $U(2)$ \cite{U(2)}, and it is connected to the fact that the first two quark families, both left- and right-handed, transform as doublets. In our model, we followed a different strategy, letting only the left-handed quarks of the first two families transform as a doublet, while the right-handed transform as singlets of $S_4$.
We manage in getting the correct up quark mass, but we ask to some parameters to combine in such a way to bring an unjustified factor of order $1/\theta_c$ or, from an alternative point of view, we ask to the phase $\zeta_q$ to be close to $\pi$. From this point of view, we only partially overcome to the problems of the previous models based on $U(2)$ and $T'$.

In the lepton sector, our model predicts a neutrino mass matrix which can be diagonalized by the TB pattern and a realistic charged lepton mass hierarchy. With respect to the $A_4$-based models, we predict an inverted hierarchy for the Majorana neutrinos and we get some interesting bounds on $|m_{ee}|$, which dominates the $0\nu2\be$-decay, on the lightest neutrino mass and on the sum of the neutrino masses:
\beq
|m_{ee}|> 14.4\;\meV\qquad\qquad|m_{\nu_3}|>0.72\;\meV\qquad\qquad\sum_i|m_{\nu i}|>89.4\;\meV\;.
\eeq
Moreover we have a prediction for $|m_{ee}|$ in function of $\De m^2_{atm}$, $r$ and the phase $\zeta$, defined in section 2,
\beq
|m_{ee}|^2=\dfrac{1}{36}\left[-(1+r)\De
m^2_{atm}+\sqrt{(\De
m^2_{atm})^2\cos^2\zeta\left(3(r-1)^2+(r+1)^2\cos^2\zeta\right)}\sin^2\zeta\right]\;.\nn
\eeq
Our predictions are quite close to the future experimental sensitivity, which are expected to reach the values of $0.090\textrm{ eV}$\cite{gerda} (GERDA), $0.020\textrm{ eV}$\cite{majorana} (Majorana), $0.050\textrm{ eV}$\cite{supernemo} (SuperNEMO), $0.015\textrm{ eV}$\cite{cuore} (CUORE) and $0.024\textrm{ eV}$\cite{exo} (EXO). These aspects can distinguish our model from all the others using only observables linked to the neutrino oscillations. Furthermore, other studies, like on some lepton flavour violating precesses, can be performed in order to complete this analysis and better characterize our proposal.

%
%%%%%%%%%%%%%%%%%%%%%%%%%%%%%%%%%%%%%%%%%%%%%%%%%%% Acknowledgments  %%%%%%%%%%%%%%%%%%%%%%%%%%%%%%%%%%%%%%%%%%%%%%%%%%%%%%%%%%%%%
%

\section*{Acknowledgments}
We thank Claudia Hagedorn for useful comments and discussions. We thank Davide Meloni for useful comments on the previous version of the paper.\\
LM would like to thank the theory group of the Institut de F\'{\i}sica Corpuscular of Valencia, where this project started, for the very kind hospitality. The work of FB has been partially supported by MEC-Valencia  MEC grant FPA2008-00319/FPA, by European Commission Contracts
MRTN-CT-2004-503369 and ILIAS/N6 RII3-CT-2004-506222 and by the foundation for Fundamental Research of Matter (FOM) and the National Organization for Scientific Research (NWO). LM recognizes that this work has been partly supported by the European Commission under contract MRTN-CT-2006-035505.
The work of SM supported by MEC-Valencia  MEC grant FPA2008-00319/FPA, by European Commission Contracts
MRTN-CT-2004-503369 and ILIAS/N6 RII3-CT-2004-506222.

%
%%%%%%%%%%%%%%%%%%%%%%%%%%%%%%%%%%%%%%%%%%%%%%%%%%% Appendix A  %%%%%%%%%%%%%%%%%%%%%%%%%%%%%%%%%%%%%%%%%%%%%%%%%%%%%%%%%%%%%
%

\newpage
\appendixA{Appendix A: The Group $S_4$}
The character table of the group $S_4$ is
\begin{table}[h]
\begin{center}
\begin{tabular}{|c|c|c|c|c|c|c|c||c|}
  \hline
  & n & h & $\chi_1$ & $\chi_{1^\prime}$ & $\chi_2$ & $\chi_3$ & $\chi_{3^\prime}$ & Example \\
  $C_1$ & 1 & 1 & 1 & 1 & 2 & 3 & 3 & \unity \\
  $C_2$ & 3 & 2 & 1 & 1 & 2 & -1 & -1 & $S^2$ \\
  $C_3$ & 8 & 3 & 1 & 1 & -1 & 0 & 0 & $T$ \\
  $C_4$ & 6 & 2 & 1 & -1 & 0 & 1 & -1 & $ST^2$ \\
  $C_5$ & 6 & 4 & 1 & -1 & 0 & -1 & 1 & $S$ \\
  \hline
\end{tabular}
\end{center}
\begin{center}
\begin{minipage}[t]{12cm}
\caption[]{Character table of $S_4$. $C_i$ are the conjugacy classes, n the number of elements in each class, h the smallest value for which $\chi^h=\unity$. In the last column we have reported an example of the elements for each class.}
\end{minipage}
\end{center}
\end{table}\\
The generators, $S$ and $T$, obey to the following rules
\beq
S^4= T^3=  (ST^2)^2=\unity
\eeq
and can be written in the different representations as
\begin{description}
  \item[representation $1_1$:] $S=1$, $T=1$
  \item[representation $1_2$:] $S=-1$, $T=1$
  \item[representation $2$:] $S=\left(
                               \begin{array}{cc}
                                 0 & 1 \\
                                 1 & 0 \\
                               \end{array}
                             \right)$, $T=\left(
                                            \begin{array}{cc}
                                              \om & 0 \\
                                              0 & \om^2 \\
                                            \end{array}
                                          \right)$
  \item[representation $3_1$:] $S=\dfrac{1}{3}\left(
                                 \begin{array}{ccc}
                                   -1 & 2\om & 2\om^2 \\
                                   2\om & 2\om^2 & -1 \\
                                   2\om^2 & -1 & 2\om \\
                                 \end{array}
                               \right)$, $T=\left(
                                              \begin{array}{ccc}
                                                1 & 0 & 0 \\
                                                0 & \om^2 & 0 \\
                                                0 & 0 & \om \\
                                              \end{array}
                                            \right)$
  \item[representation $3_2$:] $S=\dfrac{1}{3}\left(
                                 \begin{array}{ccc}
                                   1 & -2\om & -2\om^2 \\
                                   -2\om & -2\om^2 & 1 \\
                                   -2\om^2 & 1 & -2\om \\
                                 \end{array}
                               \right)$, $T=\left(
                                              \begin{array}{ccc}
                                                1 & 0 & 0 \\
                                                0 & \om^2 & 0 \\
                                                0 & 0 & \om \\
                                              \end{array}
                                            \right)$\;.
\end{description}
The 24 elements of the group belong to five conjugacy classes
\begin{center}
\begin{tabular}{ll}
  $\cC_1$ : & $\unity$ \\
  $\cC_2$ : & $S^2$, $TS^2T^2$, $S^2TS^2T^2$ \\
  $\cC_3$ : & $T$, $T^2$, $S^2T$, $S^2T^2$, $STST^2$, $STS$, $S^2TS^2$, $S^3TS$ \\
  $\cC_4$ : & $ST^2$, $T^2S$, $TST$, $TSTS^2$, $STS^2$, $S^2TS$ \\
  $\cC_5$ : & $S$, $TST^2$, $ST$, $TS$, $S^3$, $S^3T^2$\;.
\end{tabular}
\end{center}
In the 2-dimensional representation the elements are
\begin{center}
\begin{tabular}{ll}
  $\cC_{1,2}$ : & $\left(
                          \begin{array}{cc}
                            1 & 0 \\
                            0 & 1 \\
                          \end{array}
                        \right)$\;,\\
  \\
  $\cC_3$ : & $\left(
                      \begin{array}{cc}
                        \om & 0 \\
                        0 & \om^2 \\
                      \end{array}
                    \right)\;,\;
                    \left(
                      \begin{array}{cc}
                        \om^2 & 0 \\
                        0 & \om \\
                      \end{array}
                    \right)$\;,\\
  \\
  $\cC_{4,5}$ : & $\left(
                      \begin{array}{cc}
                        0 & 1 \\
                        1 & 0 \\
                      \end{array}
                    \right)\;,\;
                    \left(
                      \begin{array}{cc}
                        0 & \om \\
                        \om^2 & 0 \\
                      \end{array}
                    \right)\;,\;
                    \left(
                      \begin{array}{cc}
                        0 & \om^2 \\
                        \om & 0 \\
                      \end{array}
                    \right)\;,$
\end{tabular}
\end{center}
while for the 3-dimensional representation $3_1$ the elements are
\begin{center}
\begin{tabular}{ll}
  $\cC_1$ : & $\left(
                 \begin{array}{ccc}
                   1 & 0 & 0 \\
                   0 & 1 & 0 \\
                   0 & 0 & 1 \\
                 \end{array}
               \right)$\;, \\
  \\
  $\cC_2$ : & $\dfrac{1}{3}\left(
                 \begin{array}{ccc}
                   -1 & 2 & 2 \\
                   2 & -1 & 2 \\
                   2 & 2 & -1 \\
                 \end{array}
               \right)\;,\;
               \dfrac{1}{3}\left(
                 \begin{array}{ccc}
                   -1 & 2\om & 2\om^2 \\
                   2\om^2 & -1 & 2\om \\
                   2\om & 2\om^2 & -1 \\
                 \end{array}
               \right)\;,\;
               \dfrac{1}{3}\left(
                 \begin{array}{ccc}
                   -1 & 2\om^2 & 2\om \\
                   2\om & -1 & 2\om^2 \\
                   2\om^2 & 2\om & -1 \\
                 \end{array}
               \right)$\;, \\
  \\
  $\cC_3$ : & $\left(
                 \begin{array}{ccc}
                   1 & 0 & 0 \\
                   0 & \om^2 & 0 \\
                   0 & 0 & \om \\
                 \end{array}
               \right)\;,\;
               \left(
                 \begin{array}{ccc}
                   1 & 0 & 0 \\
                   0 & \om & 0 \\
                   0 & 0 & \om^2 \\
                 \end{array}
               \right)$\;,\\[0.6cm]
             & $\dfrac{1}{3}\left(
                 \begin{array}{ccc}
                   -1 & 2\om^2 & 2\om \\
                   2 & -\om^2 & 2\om \\
                   2 & 2\om^2 & -\om \\
                 \end{array}
               \right)\;,\;
               \dfrac{1}{3}\left(
                 \begin{array}{ccc}
                   -1 & 2\om & 2\om^2 \\
                   2 & -\om & 2\om^2 \\
                   2 & 2\om & -\om^2 \\
                 \end{array}
               \right)\;,\;
               \dfrac{1}{3}\left(
                 \begin{array}{ccc}
                   -1 & 2 & 2 \\
                   2\om^2 & -\om^2 & 2\om^2 \\
                   2\om & 2\om & -\om \\
                 \end{array}
               \right)$\;,\\[0.6cm]
             & $\dfrac{1}{3}\left(
                 \begin{array}{ccc}
                   -1 & 2\om^2 & 2\om \\
                   2\om^2 & -\om & 2 \\
                   2\om & 2 & -\om^2 \\
                 \end{array}
               \right)\;,\;
               \dfrac{1}{3}\left(
                 \begin{array}{ccc}
                   -1 & 2\om & 2\om^2 \\
                   2\om & -\om^2 & 2 \\
                   2\om^2 & 2 & -\om \\
                 \end{array}
               \right)\;,\;
               \dfrac{1}{3}\left(
                 \begin{array}{ccc}
                   -1 & 2 & 2 \\
                   2\om & -\om & 2\om \\
                   2\om^2 & 2\om^2 & -\om^2 \\
                 \end{array}
               \right)$\;,\\
  \\
  $\cC_4$ : & $\dfrac{1}{3}\left(
                 \begin{array}{ccc}
                   -1 & 2\om^2 & 2\om \\
                   2\om & 2 & -\om^2 \\
                   2\om^2 & -\om & 2 \\
                 \end{array}
               \right)\;,\;
               \dfrac{1}{3}\left(
                 \begin{array}{ccc}
                   -1 & 2\om & 2\om^2 \\
                   2\om^2 & 2 & -\om \\
                   2\om & -\om^2 & 2 \\
                 \end{array}
               \right)\;,\;
               \dfrac{1}{3}\left(
                 \begin{array}{ccc}
                   -1 & 2 & 2 \\
                   2 & 2 & -1 \\
                   2 & -1 & 2 \\
                 \end{array}
               \right)$\;,\\[0.6cm]
            &  $\left(
                 \begin{array}{ccc}
                   1 & 0 & 0 \\
                   0 & 0 & 1 \\
                   0 & 1 & 0 \\
                 \end{array}
               \right)\;,\;
               \left(
                 \begin{array}{ccc}
                   1 & 0 & 0 \\
                   0 & 0 & \om \\
                   0 & \om^2 & 0 \\
                 \end{array}
               \right)\;,\;
               \left(
                 \begin{array}{ccc}
                   1 & 0 & 0 \\
                   0 & 0 & \om^2 \\
                   0 & \om & 0 \\
                 \end{array}
               \right)$\;,
\end{tabular}
\end{center}
\begin{center}
\begin{tabular}{ll}
  $\cC_5$ : & $\dfrac{1}{3}\left(
                 \begin{array}{ccc}
                   -1 & 2\om & 2\om^2 \\
                   2\om & 2\om^2 & -1 \\
                   2\om^2 & -1 & 2\om \\
                 \end{array}
               \right)\;,\;
               \dfrac{1}{3}\left(
                 \begin{array}{ccc}
                   -1 & 2\om^2 & 2\om \\
                   2 & 2\om^2 & -\om \\
                   2 & -\om^2 & 2\om \\
                 \end{array}
               \right)\;,\;
               \dfrac{1}{3}\left(
                 \begin{array}{ccc}
                   -1 & 2 & 2 \\
                   2\om & 2\om & -\om \\
                   2\om^2 & -\om^2 & 2\om^2 \\
                 \end{array}
               \right)$\;,\\[0.6cm]
            &  $\dfrac{1}{3}\left(
                 \begin{array}{ccc}
                   -1 & 2\om & 2\om^2 \\
                   2 & 2\om & -\om^2 \\
                   2 & -\om & 2\om^2 \\
                 \end{array}
               \right)\;,\;
               \dfrac{1}{3}\left(
                 \begin{array}{ccc}
                   -1 & 2\om^2 & 2\om \\
                   2\om^2 & 2\om & -1 \\
                   2\om & -1 & 2\om^2 \\
                 \end{array}
               \right)\;,\;
              \dfrac{1}{3}\left(
                 \begin{array}{ccc}
                   -1 & 2 & 2 \\
                   2\om^2 & 2\om^2 & -\om^2 \\
                   2\om & -\om & 2\om \\
                 \end{array}
               \right)$\;,
\end{tabular}
\end{center}
and finally for the 3-dimensional representation $3_2$, the matrices representing the elements of the group can be found from those just listed for the representation $3_1$: for $\cC_{1,2,3}$ are the same, while for $\cC_{4,5}$ are the opposite. It is connected with the generator $S$, which changes sign in the $3_1$ and $3_2$ representations: the elements in $\cC_{1,2,3}$ contain an even number of $S$, while those in $\cC_{4,5}$ contain an odd number of it.\\
\\
We now report the Clebsch-Gordan coefficients for our basis. In the following we
use $\alpha_i$ to indicate the elements of the first
representation of the product and $\beta_i$ to indicate
those of the second representation.\\
We start with all the
multiplication rules which include the 1-dimensional
representations:
\[
\begin{array}{lcl}
1_1\otimes\eta&=&\eta\otimes1_1=\eta\quad\text{with $\eta$ any
representation}\\[-10pt]
\\[8pt]
1_2\otimes1_2&=&1_1\sim\alpha\beta\\[-10pt]
\\[8pt]
1_2\otimes2&=&2\sim\left(\begin{array}{c}
                    \alpha\beta_1 \\
                    -\alpha\beta_2 \\
            \end{array}\right)\\[-10pt]
\\[8pt]
1_2\otimes3_1&=&3_2\sim\left(\begin{array}{c}
                    \alpha\beta_1 \\
                    \alpha\beta_2 \\
                    \alpha\beta_3 \\
                    \end{array}\right)\\[-10pt]
\\[8pt]
1_2\otimes3_2&=&3_1\sim\left(\begin{array}{c}
                            \alpha\beta_1 \\
                            \alpha\beta_2 \\
                            \alpha\beta_3 \\
                    \end{array}\right)
\end{array}
\]
The multiplication rules with the 2-dimensional
representation are the following:
\[
\begin{array}{ll}
2\otimes2=1_1\oplus1_2\oplus2&\quad
\text{with}\quad\left\{\begin{array}{l}
                    1_1\sim\alpha_1\beta_2+\alpha_2\beta_1\\[-10pt]
                    \\[8pt]
                    1_2\sim\alpha_1\beta_2-\alpha_2\beta_1\\[-10pt]
                    \\[8pt]
                    2\sim\left(\begin{array}{c}
                        \alpha_2\beta_2 \\
                        \alpha_1\beta_1 \\
                    \end{array}\right)
                    \end{array}
            \right.\\[-10pt]
\\[8pt]
2\otimes3_1=3_1\oplus3_2&\quad
\text{with}\quad\left\{\begin{array}{l}
                    3_1\sim\left(\begin{array}{c}
                        \alpha_1\beta_2+\alpha_2\beta_3 \\
                        \alpha_1\beta_3+\alpha_2\beta_1 \\
                        \alpha_1\beta_1+\alpha_2\beta_2 \\
                    \end{array}\right)\\[-10pt]
                    \\[8pt]
                    3_2\sim\left(\begin{array}{c}
                        \alpha_1\beta_2-\alpha_2\beta_3\\
                        \alpha_1\beta_3-\alpha_2\beta_1 \\
                        \alpha_1\beta_1-\alpha_2\beta_2 \\
                    \end{array}\right)\\
                    \end{array}
            \right.\\[-10pt]
\\[8pt]
2\otimes3_2=3_1\oplus3_2&\quad
\text{with}\quad\left\{\begin{array}{l}
                    3_1\sim\left(\begin{array}{c}
                        \alpha_1\beta_2-\alpha_2\beta_3\\
                        \alpha_1\beta_3-\alpha_2\beta_1 \\
                        \alpha_1\beta_1-\alpha_2\beta_2 \\
                    \end{array}\right)\\[-10pt]
                    \\[8pt]
                    3_2\sim\left(\begin{array}{c}
                        \alpha_1\beta_2+\alpha_2\beta_3 \\
                        \alpha_1\beta_3+\alpha_2\beta_1 \\
                        \alpha_1\beta_1+\alpha_2\beta_2 \\
                    \end{array}\right)\\
                    \end{array}
            \right.\\
\end{array}
\]
The multiplication rules with the 3-dimensional
representations are the following:
\[
\begin{array}{ll}
3_1\otimes3_1=3_2\otimes3_2=1_1\oplus2\oplus3_1\oplus3_2\qquad
\text{with}\quad\left\{
\begin{array}{l}
1_1\sim\alpha_1\beta_1+\alpha_2\beta_3+\alpha_3\beta_2 \\[-10pt]
                    \\[8pt]
2\sim\left(
     \begin{array}{c}
       \al_2\be_2+\al_1\be_3+\al_3\be_1 \\
       \al_3\be_3+\al_1\be_2+\al_2\be_1 \\
     \end{array}
   \right)\\[-10pt]
   \\[8pt]
3_1\sim\left(\begin{array}{c}
         2\al_1\be_1-\alpha_2\beta_3-\alpha_3\beta_2 \\
         2\al_3\be_3-\alpha_1\beta_2-\alpha_2\beta_1 \\
         2\al_2\be_2-\alpha_1\beta_3-\alpha_3\beta_1 \\
        \end{array}\right)\\[-10pt]
        \\[8pt]
3_2\sim\left(\begin{array}{c}
         \alpha_2\beta_3-\alpha_3\beta_2 \\
         \alpha_1\beta_2-\alpha_2\beta_1 \\
         \alpha_3\beta_1-\alpha_1\beta_3 \\
    \end{array}\right)
\end{array}\right.
\end{array}
\]
\[
\begin{array}{ll}
3_1\otimes3_2=1_2\oplus2\oplus3_1\oplus3_2\qquad
\text{with}\quad\left\{
\begin{array}{l}
1_2\sim\alpha_1\beta_1+\alpha_2\beta_3+\alpha_3\beta_2\\[-10pt]
        \\[8pt]
2\sim\left(
     \begin{array}{c}
       \al_2\be_2+\al_1\be_3+\al_3\be_1 \\
       -\al_3\be_3-\al_1\be_2-\al_2\be_1 \\
     \end{array}
   \right)\\[-10pt]
        \\[8pt]
3_1\sim\left(\begin{array}{c}
         \alpha_2\beta_3-\alpha_3\beta_2 \\
         \alpha_1\beta_2-\alpha_2\beta_1 \\
         \alpha_3\beta_1-\alpha_1\beta_3 \\
    \end{array}\right)\\[-10pt]
        \\[8pt]
3_2\sim\left(\begin{array}{c}
         2\al_1\be_1-\alpha_2\beta_3-\alpha_3\beta_2 \\
         2\al_3\be_3-\alpha_1\beta_2-\alpha_2\beta_1 \\
         2\al_2\be_2-\alpha_1\beta_3-\alpha_3\beta_1 \\
    \end{array}\right)\\
\end{array}\right.
\end{array}
\]

%
%%%%%%%%%%%%%%%%%%%%%%%%%%%%%%%%%%%%%%%%%%%%%%%%%%% Appendix B  %%%%%%%%%%%%%%%%%%%%%%%%%%%%%%%%%%%%%%%%%%%%%%%%%%%%%%%%%%%%%
%

\appendixB{Appendix B: The Vacuum Alignment at NLO}
In this section there is the analysis for the corrections to the vacuum alignment introduced by the higher dimensional operators.
In table \ref{AppendixB:table:flavon_transformation} there is a summary of the transformation properties of the flavons and of the driving fields.
\begin{table}[h]
\begin{center}
\begin{tabular}{|c||c|c|c|c||c|c|c||c|c|}
  \hline
  &&&&&&&&& \\[-0,3cm]
  & $\Delta$ & $\vphi$ & $\Delta^0$ & $\vphi^0$ & $\psi$ & $\eta$ & $\psi^0$ & $\xi'$ & $\xi'^0$ \\
  &&&&&&&&& \\[-0,3cm]
  \hline
  &&&&&&&&& \\[-0,3cm]
  $S_4$ & $3_1$ & 2 & $3_2$ & 2 & $3_1$ & $2$ & $3_1$ & $1_2$ & $1_2$ \\
  &&&&&&&&& \\[-0,3cm]
  $Z_5$ & $\om^3$ & $\om^3$ & $\om^4$ & $\om^4$ & $\om^2$ & $\om^2$ & $\om$ & 1 & 1 \\
  \hline
  \end{tabular}
\end{center}
\begin{center}
\begin{minipage}[t]{12cm}
\caption[]{Transformation properties of the flavons and the driving fields.}
\label{AppendixB:table:flavon_transformation}
\end{minipage}
\end{center}
\end{table}

The part of the superpotential depending on the driving fields $\De^0$, $\vphi^0$, $\psi^0$ and $\xi'^0$ is modified into
\beq
w_d=w^0_d+\delta w_d\;.
\eeq
The leading order contribution is
\bac{rcl}
w_d&=&g_1(\De^0\De\vphi)+g_2(\vphi^0\De\De)+g_3(\vphi^0\vphi\vphi)+\\[0.3cm]
&&+f_1(\psi^0\psi\psi)+f_2(\psi^0\psi\eta)+\\[0.3cm]
&&+M_{\xi'}\xi'^0\xi'+h_1\xi'^0(\eta\vphi)'
\eac
and the minimum is
\bac{ll}
\mean{\De}\sim\left(
                     \begin{array}{c}
                       1 \\
                       1 \\
                       1 \\
                     \end{array}
                   \right)v_\De\qquad
\qquad&\mean{\vphi}\sim\left(
                     \begin{array}{c}
                       1 \\
                       1 \\
                     \end{array}
                   \right)v_\vphi\\
\\
\mean{\psi}\sim\left(
                     \begin{array}{c}
                       0 \\
                       1 \\
                       0 \\
                     \end{array}
                   \right)v_\psi\qquad
\qquad&\mean{\eta}\sim\left(
                     \begin{array}{c}
                       0 \\
                       1 \\
                     \end{array}
                   \right)v_\eta\\
\\
\mean{\xi'}\sim v_{\xi'}
\label{Fvev:NLO:LO}
\eac
where
\beq
v_\De^2=-\dfrac{g_3}{3g_2}v_\vphi^2\qquad\qquad v_\psi=-\dfrac{f_2}{2f_1}v_\eta\qquad v_{xi'}=\dfrac{h_1}{M_{\xi'}}v_\eta v_\vphi\;.
\label{Fvev:NLO:NLO_parameters}
\eeq
The remaining part, $\delta w_d$, is the most general quartic, $S_4$-invariant polynomial linear in the driving fields:
\beq
\delta w_d=\dfrac{1}{\La}\left(\sum_{i=1}^5x_iI_i^{\De^0}+\sum_{i=1}^{6}w_iI_i^{\vphi^0}+\sum_{i=1}^7s_iI_i^{\psi^0}+
\sum_{i=1}^2v_iI_i^{\xi'^0}\right)
\eeq
where $x_i$, $w_i$, $s_i$ and $v_i$ are coefficients and $\left\{I_i^{\De^0},\;I_i^{\vphi^0},\;I_i^{\psi^0},\;I_i^{\xi'^0}\right\}$ represents a basis of independent quartic invariants:
\bac{ll}
I_1^{\De^0}=(\De^0(\De\vphi)_{3_1})'\xi'\qquad\qquad&
I_4^{\De^0}=((\De^0\eta)_{3_1}(\psi\psi)_{3_1})\\
I_2^{\De^0}=(\De^0(\De\De)_{3_1})'\xi'\qquad\qquad&
I_5^{\De^0}=((\De^0\psi)_2(\eta\eta)_2)\\
I_3^{\De^0}=((\De^0\psi)_2(\psi\psi)_2)&\\
\\
I_1^{\vphi^0}=(\vphi^0(\De\De)_2)'\xi'\qquad\qquad&
I_4^{\vphi^0}=(\vphi^0\eta)(\psi\psi)\\
I_2^{\vphi^0}=(\vphi^0(\vphi\vphi)_2)'\xi'\qquad\qquad&
I_5^{\vphi^0}=(\vphi^0\eta)(\eta\eta)\\
I_3^{\vphi^0}=((\vphi^0\eta)_2(\psi\psi)_2)\qquad\qquad&\\
\\
I_1^{\psi^0}=((\psi^0\psi)_2\eta)'\xi'\qquad\qquad&
I_4^{\psi^0}=((\psi^0\vphi)_{3_1}(\De\De)_{3_1})\\
I_2^{\psi^0}=((\psi^0\De)_2(\De\De)_2)\qquad\qquad&
I_5^{\psi^0}=((\psi^0\De)_2(\vphi\vphi)_2)\\
I_3^{\psi^0}=(\psi^0\De)(\De\De)\qquad\qquad&
I_6^{\psi^0}=(\psi^0\De)(\vphi\vphi)\\
\\
I_1^{\xi'^0}=\xi'^0\xi'\xi'\xi'\qquad\qquad&
I_2^{\xi'^0}=\xi'^0\xi'(\vphi\eta)\\
I_2^{\xi'^0}=\xi'^0\xi'(\De\psi)\qquad\qquad&\\
\eac
The new minimum for $\De$, $\vphi$, $\psi$, $\eta$ and $\xi'$ is obtained by searching for the zeros of the F terms, the first derivative of $w_d+\delta w_d$, associated to the driving fields $\De^0$, $\vphi^0$, $\psi^0$ and $\xi'^0$. We look for a solution that perturbs eq.(\ref{Fvev:NLO:LO}) to first order in the $1/\La$ expansion: denoting the general flavon field with $\Phi$, we can write the new VEVs as
\beq
\mean{\Phi_i}=\mean{\Phi_i}^{(LO)}+\de\Phi_i\;.
\eeq
The minimum conditions become equations in the unknown $\de\Phi_i$, $v_\vphi$ and $v_\eta$. By keeping only the first order in the expansion, we see that the equations can be separated into different groups: the first five concern only the neutrino sector, the second three only the charged lepton one and the last one connects the two sectors. Finally all the perturbations are non vanishing, a part $\de\eta_1$ and $\de\eta_2$ and one of the perturbations in the neutrino sector, which remain undetermined. On the other hand the NLO terms fixes the relation between $v_\vphi$ and $v_\eta$. We can conclude that the VEV alignment in eq.(\ref{Fvev:NLO:LO}) is stable under the NLO corrections and the deviations are of relative order $u$ with respect to the LO results.

%
%%%%%%%%%%%%%%%%%%%%%%%%%%%%%%%%%%%%%%%%%%%%%%%%% Bibliography %%%%%%%%%%%%%%%%%%%%%%%%%%%%%%%%%%%%%%%%%%%%%%%%%%%%%%%%%%%%%%%
%


\begin{thebibliography}{99}

\bibitem{Data}
A.~Strumia and F.~Vissani,
%  {\it Neutrino masses and mixings and.},
  arXiv:hep-ph/0606054;
  %%CITATION = HEP-PH/0606054;%%
G.~L.~Fogli {\it et al.},
%  {\it Neutrino mass and mixing: 2006 status},
  Nucl.\ Phys.\ Proc.\ Suppl.\  {\bf 168} (2007) 341;
  %%CITATION = NUPHZ,168,341;%%
M.~C.~Gonzalez-Garcia and M.~Maltoni,
%  {\it Phenomenology with Massive Neutrinos},
  Phys.\ Rept.\  {\bf 460} (2008) 1
  [arXiv:0704.1800 [hep-ph]];
  %%CITATION = PRPLC,460,1;%%
T.~Schwetz,
%  {\it Neutrino oscillations: present status and outlook},
  AIP Conf.\ Proc.\  {\bf 981} (2008) 8
  [arXiv:0710.5027 [hep-ph]];
  %%CITATION = APCPC,981,8;%%
M.~C.~Gonzalez-Garcia and M.~Maltoni,
%  {\it Status of Oscillation plus Decay of Atmospheric and Long-Baseline Neutrinos},
  Phys.\ Lett.\  B {\bf 663} (2008) 405
  [arXiv:0802.3699 [hep-ph]];
  %%CITATION = PHLTA,B663,405;%%
A.~Bandyopadhyay, S.~Choubey, S.~Goswami, S.~T.~Petcov and D.~P.~Roy,
%  {\it Neutrino Oscillation Parameters After High Statistics KamLAND Results},
  arXiv:0804.4857 [hep-ph].
  %%CITATION = ARXIV:0804.4857;%%



\bibitem{Fogli:Indication}
G.~L.~Fogli, E.~Lisi, A.~Marrone, A.~Palazzo and A.~M.~Rotunno,
  %``What we (would like to) know about the neutrino mass,''
  arXiv:0809.2936 [hep-ph];
  %%CITATION = ARXIV:0809.2936;%%
G.~L.~Fogli, E.~Lisi, A.~Marrone, A.~Palazzo and A.~M.~Rotunno,
  %``Hints of theta_13>0 from global neutrino data analysis,''
  Phys.\ Rev.\ Lett.\  {\bf 101} (2008) 141801
  [arXiv:0806.2649 [hep-ph]].
  %%CITATION = PRLTA,101,141801;%%



\bibitem{Maltoni:Indication}
T.~Schwetz, M.~Tortola and J.~W.~F.~Valle,
  %``Three-flavour neutrino oscillation update,''
  New J.\ Phys.\  {\bf 10} (2008) 113011
  [arXiv:0808.2016 [hep-ph]];
  %%CITATION = NJOPF,10,113011;%%
M.~Maltoni and T.~Schwetz,
  %``Three-flavour neutrino oscillation update and comments on possible hints
  %for a non-zero theta_{13},''
  arXiv:0812.3161 [hep-ph].
  %%CITATION = ARXIV:0812.3161;%%



\bibitem{Ardellier:2006mn}
F.~Ardellier {\it et al.} [Double Chooz Collaboration],
%``Double Chooz: A search for the neutrino mixing angle theta(13),''
arXiv:hep-ex/0606025.
%%CITATION = HEP-EX/0606025;%%



\bibitem{Wang:2006ca}
Y.~f.~Wang,
%``Measuring sin**2(2theta(13)) with the Daya Bay nuclear reactors,''
arXiv:hep-ex/0610024.
%%CITATION = HEP-EX/0610024;%%



\bibitem{PereiraeSousa:2005rf}
A.~B.~Pereira e Sousa,
%``Studies of $\nu_\mu \to \nu_e$ oscillation appearance in the MINOS experiment,''
FERMILAB-THESIS-2005-67.
%%CITATION = UMI-31-99667;%%



\bibitem{TB}
P.~F.~Harrison, D.~H.~Perkins and W.~G.~Scott,
%  {\it Tri-bimaximal mixing and the neutrino oscillation data},
  Phys.\ Lett.\ B {\bf 530} (2002) 167
  [arXiv:hep-ph/0202074];
  %%CITATION = PHLTA,B530,167;%%
P.~F.~Harrison and W.~G.~Scott,
%  {\it Symmetries and generalisations of tri-bimaximal neutrino mixing},
  Phys.\ Lett.\ B {\bf 535} (2002) 163
  [arXiv:hep-ph/0203209];
  %%CITATION = PHLTA,B535,163;%%
Z.~z.~Xing,
%  {\it Nearly tri-bimaximal neutrino mixing and CP violation},
  Phys.\ Lett.\ B {\bf 533} (2002) 85
  [arXiv:hep-ph/0204049];
  %%CITATION = PHLTA,B533,85;%%
P.~F.~Harrison and W.~G.~Scott,
%  {\it Permutation symmetry, tri-bimaximal neutrino mixing and the S3 group characters},
  Phys.\ Lett.\ B {\bf 557} (2003) 76
  [arXiv:hep-ph/0302025];
  %%CITATION = PHLTA,B557,76;%%
%  {\it Status of tri- / bi-maximal neutrino mixing},
  arXiv:hep-ph/0402006;
  %%CITATION = HEP-PH/0402006;%%
%  {\it The simplest neutrino mass matrix},
  Phys.\ Lett.\  B {\bf 594} (2004) 324
  [arXiv:hep-ph/0403278].
  %%CITATION = PHLTA,B594,324;%%



\bibitem{af:extra}
G.~Altarelli and F.~Feruglio,
  %``Tri-bimaximal neutrino mixing from discrete symmetry in extra
  %dimensions,''
  Nucl.\ Phys.\  B {\bf 720} (2005) 64
  [arXiv:hep-ph/0504165].
  %%CITATION = NUPHA,B720,64;%%



\bibitem{TBA4}
E.~Ma and G.~Rajasekaran,
%  {\it Softly broken A(4) symmetry for nearly degenerate neutrino masses},
  Phys.\ Rev.\ D {\bf 64} (2001) 113012
  [arXiv:hep-ph/0106291];
K.~S.~Babu, E.~Ma and J.~W.~F.~Valle,
%  {\it Underlying A(4) symmetry for the neutrino mass matrix and the quark  mixing matrix},
  Phys.\ Lett.\ B {\bf 552} (2003) 207
  [arXiv:hep-ph/0206292];
M.~Hirsch, J.~C.~Romao, S.~Skadhauge, J.~W.~F.~Valle and A.~Villanova del Moral,
%  {\it Degenerate neutrinos from a supersymmetric A(4) model},
  arXiv:hep-ph/0312244;
%  {\it Phenomenological tests of supersymmetric A(4) family symmetry model of neutrino mass},
  Phys.\ Rev.\  D {\bf 69} (2004) 093006
  [arXiv:hep-ph/0312265];
E.~Ma,
%  {\it Quark mass matrices in the A(4) model},
  Mod.\ Phys.\ Lett.\ A {\bf 17} (2002) 627
  [arXiv:hep-ph/0203238];
  %%CITATION = HEP-PH 0203238;%%
%  {\it A(4) origin of the neutrino mass matrix},
  Phys.\ Rev.\ D {\bf 70} (2004) 031901
  [arXiv:hep-ph/0404199];
%  {\it Non-Abelian discrete symmetries and neutrino masses: Two examples},
  New J.\ Phys.\  {\bf 6} (2004) 104
  [arXiv:hep-ph/0405152];
%  {\it Non-Abelian discrete family symmetries of leptons and quarks},
  arXiv:hep-ph/0409075;
  %%CITATION = HEP-PH/0409075;%%
%  {\it Aspects of the tetrahedral neutrino mass matrix},
  Phys.\ Rev.\  D {\bf 72} (2005) 037301
  [arXiv:hep-ph/0505209];
%  {\it Tetrahedral family symmetry and the neutrino mixing matrix},
  Mod.\ Phys.\ Lett.\ A {\bf 20} (2005) 2601
  [arXiv:hep-ph/0508099];
%  {\it Tribimaximal neutrino mixing from a supersymmetric model with A4 family symmetry},
  Phys.\ Rev.\  D {\bf 73} (2006) 057304
  [arXiv:hep-ph/0511133];
%  {\it Suitability of A(4) as a family symmetry in grand unification},
  Mod.\ Phys.\ Lett.\  A {\bf 21} (2006) 2931
  [arXiv:hep-ph/0607190];
%  {\it Supersymmetric A(4) x Z(3) and A(4) realizations of neutrino  tribimaximal mixing without and with corrections},
  Mod.\ Phys.\ Lett.\  A {\bf 22} (2007) 101
  [arXiv:hep-ph/0610342];
S.~L.~Chen, M.~Frigerio and E.~Ma,
%  {\it Hybrid seesaw neutrino masses with A(4) family symmetry},
  Nucl.\ Phys.\  B {\bf 724} (2005) 423
  [arXiv:hep-ph/0504181];
  %%CITATION = NUPHA,B724,423;%%
K.~S.~Babu and X.~G.~He,
%  {\it Model of geometric neutrino mixing},
  arXiv:hep-ph/0507217;
A.~Zee,
% {\it Obtaining the neutrino mixing matrix with the tetrahedral group},
  Phys.\ Lett.\ B {\bf 630} (2005) 58
  [arXiv:hep-ph/0508278];
X.~G.~He, Y.~Y.~Keum and R.~R.~Volkas,
%  {\it A(4) flavour symmetry breaking scheme for understanding quark and  neutrino mixing angles},
  JHEP {\bf 0604} (2006) 039
  [arXiv:hep-ph/0601001];
B.~Adhikary, B.~Brahmachari, A.~Ghosal, E.~Ma and M.~K.~Parida,
%  {\it A(4) symmetry and prediction of U(e3) in a modified Altarelli-Feruglio model},
  Phys.\ Lett.\ B {\bf 638} (2006) 345
  [arXiv:hep-ph/0603059];
L.~Lavoura and H.~Kuhbock,
%  {\it Predictions of an A(4) model with a five-parameter neutrino mass matrix},
  Mod.\ Phys.\ Lett.\  A {\bf 22} (2007) 181
  [arXiv:hep-ph/0610050];
S.~F.~King and M.~Malinsky,
%  {\it A(4) family symmetry and quark-lepton unification},
  Phys.\ Lett.\  B {\bf 645} (2007) 351
  [arXiv:hep-ph/0610250];
S.~Morisi, M.~Picariello and E.~Torrente-Lujan,
%  {\it A model for fermion masses and lepton mixing in SO(10) x A(4)},
  Phys.\ Rev.\  D {\bf 75} (2007) 075015
  [arXiv:hep-ph/0702034];
F.~Yin,
% {\it Neutrino mixing matrix in the 3-3-1 model with heavy leptons and A(4) symmetry},
  Phys.\ Rev.\  D {\bf 75} (2007) 073010
  [arXiv:0704.3827 [hep-ph]];
F.~Bazzocchi, S.~Morisi and M.~Picariello,
%  {\it Embedding A(4) into left-right flavor symmetry: Tribimaximal neutrino mixing and fermion hierarchy},
  Phys.\ Lett.\  B {\bf 659} (2008) 628
  [arXiv:0710.2928 [hep-ph]];
M.~Honda and M.~Tanimoto,
%  {\it Deviation from tri-bimaximal neutrino mixing in A(4) flavor symmetry},
  Prog.\ Theor.\ Phys.\  {\bf 119} (2008) 583
  [arXiv:0801.0181 [hep-ph]];
B.~Brahmachari, S.~Choubey and M.~Mitra,
%  {\it The A(4) flavor symmetry and neutrino phenomenology},
  Phys.\ Rev.\  D {\bf 77} (2008) 073008
  [Erratum-ibid.\  D {\bf 77} (2008) 119901]
  [arXiv:0801.3554 [hep-ph]];
F.~Bazzocchi, S.~Morisi, M.~Picariello and E.~Torrente-Lujan,
  %``Embedding A4 into SU(3)xU(1) flavor symmetry: Large neutrino mixing and
  %fermion mass hierarchy in SO(10) GUT,''
  J.\ Phys.\ G {\bf 36} (2009) 015002
  [arXiv:0802.1693 [hep-ph]];
  %%CITATION = JPHGB,G36,015002;%%
B.~Adhikary and A.~Ghosal,
  %``Nonzero U_{e3}, CP violation and leptogenesis in a see-saw type softly
  %broken A_4 symmetric model,''
  Phys.\ Rev.\  D {\bf 78} (2008) 073007
  [arXiv:0803.3582 [hep-ph]];
  %%CITATION = PHRVA,D78,073007;%%
P.~H.~Frampton and S.~Matsuzaki,
  %``Renormalizable $A_4$ Model for Lepton Sector,''
  arXiv:0806.4592 [hep-ph];
  %%CITATION = ARXIV:0806.4592;%%
F.~Bazzocchi, M.~Frigerio and S.~Morisi,
  %``Fermion masses and mixing in models with SO(10) x A_4 symmetry,''
  arXiv:0809.3573 [hep-ph];
  %%CITATION = ARXIV:0809.3573;%%
S.~Baek and M.~C.~Oh,
  %``Neutrino mass matrix in triplet Higgs models with $A_4$ symmetry,''
  arXiv:0812.2704 [hep-ph].
  %%CITATION = ARXIV:0812.2704;%%



\bibitem{af:modular}
G.~Altarelli and F.~Feruglio,
  %``Tri-Bimaximal Neutrino Mixing, A4 and the Modular Symmetry,''
  Nucl.\ Phys.\  B {\bf 741} (2006) 215
  [arXiv:hep-ph/0512103].
  %%CITATION = NUPHA,B741,215;%%



\bibitem{afl}
G.~Altarelli, F.~Feruglio and Y.~Lin,
% {\it Tri-bimaximal neutrino mixing from orbifolding},
  Nucl.\ Phys.\  B {\bf 775} (2007) 31
  [arXiv:hep-ph/0610165].
  %%CITATION = NUPHA,B775,31;%%



\bibitem{bkm}
  F.~Bazzocchi, S.~Kaneko and S.~Morisi,
  %``A SUSY A4 model for fermion masses and mixings,''
  JHEP {\bf 0803} (2008) 063
  [arXiv:0707.3032 [hep-ph]].
  %%CITATION = JHEPA,0803,063;%%



\bibitem{afh}
  G.~Altarelli, F.~Feruglio and C.~Hagedorn,
  %``A SUSY SU(5) Grand Unified Model of Tri-Bimaximal Mixing from A4,''
  JHEP {\bf 0803} (2008) 052
  [arXiv:0802.0090 [hep-ph]].
  %%CITATION = JHEPA,0803,052;%%



\bibitem{linyin}
Y.~Lin,
  %``A predictive A4 model, Charged Lepton Hierarchy and Tri-bimaximal Sum
  %Rule,''
  arXiv:0804.2867 [hep-ph].
  %%CITATION = ARXIV:0804.2867;%%



\bibitem{hmv}
  M.~Hirsch, S.~Morisi and J.~W.~F.~Valle,
  %``Tri-bimaximal neutrino mixing and neutrinoless double beta decay,''
  Phys.\ Rev.\  D {\bf 78} (2008) 093007
  [arXiv:0804.1521 [hep-ph]].
  %%CITATION = PHRVA,D78,093007;%%


\bibitem{bfm}
F.~Bazzocchi, M.~Frigerio and S.~Morisi,
  %``Fermion masses and mixing in models with SO(10) x A_4 symmetry,''
  arXiv:0809.3573 [hep-ph];
  %%CITATION = ARXIV:0809.3573;%%



\bibitem{Grimus:2008tt}
  W.~Grimus and L.~Lavoura,
  %``A model for trimaximal lepton mixing,''
  JHEP {\bf 0809}, 106 (2008)
  [arXiv:0809.0226 [hep-ph]].
  %%CITATION = JHEPA,0809,106;%%



\bibitem{fhlm:LFV}
F.~Feruglio, C.~Hagedorn, Y.~Lin and L.~Merlo,
  %``Lepton Flavour Violation in Models with A4 Flavour Symmetry,''
  arXiv:0807.3160 [hep-ph];
  %%CITATION = ARXIV:0807.3160;%%



\bibitem{U(2)}
R.~Barbieri, G.~R.~Dvali and L.~J.~Hall,
    %``Predictions From A $\Ud$ Flavour Symmetry In
    %Supersymmetric Theories'',
    Phys.\ Lett.\ B {\bf 377}, 76 (1996)
    [arXiv:hep-ph/9512388];
    %%CITATION = HEP-PH 9512388;%%
R.~Barbieri, L.~J.~Hall, S.~Raby and A.~Romanino,
    %``Unified Theories With $\Ud$ Flavor Symmetry'',
    Nucl.\ Phys.\ B {\bf 493}, 3 (1997)
    [arXiv:hep-ph/9610449];
    %%CITATION = HEP-PH 9610449;%%
R.~Barbieri, L.~J.~Hall and A.~Romanino,
    %``Consequences Of A $\Ud$ Flavour Symmetry'',
    Phys.\ Lett.\ B {\bf 401}, 47 (1997)
    [arXiv:hep-ph/9702315];
    %%CITATION = HEP-PH 9702315;%%
A.~Aranda, C.~D.~Carone and R.~F.~Lebed,
    %``$\Ud$ Flavor Physics Without $\Ud$ Symmetry'',
    Phys.\ Lett.\ B {\bf 474}, 170 (2000)
    [arXiv:hep-ph/9910392];
    %%CITATION = HEP-PH 9910392;%%
A.~Aranda, C.~D.~Carone and R.~F.~Lebed,
    %``Maximal Neutrino Mixing From A Minimal Flavor Symmetry'',
    Phys.\ Rev.\ D {\bf 62}, 016009 (2000)
    [arXiv:hep-ph/0002044].
    %%CITATION = HEP-PH 0002044;%%



\bibitem{fhlm:Tp}
F.~Feruglio, C.~Hagedorn, Y.~Lin and L.~Merlo,
% {\it Tri-bimaximal neutrino mixing and quark masses from a discrete flavour symmetry},
  Nucl.\ Phys.\  B {\bf 775} (2007) 120
  [arXiv:hep-ph/0702194].
  %%CITATION = NUPHA,B775,120;%%



\bibitem{Tp}
M.~C.~Chen and K.~T.~Mahanthappa,
%  {\it CKM and Tri-bimaximal MNS Matrices in a $SU(5) \times ^{(d)}T$ Model},
  Phys.\ Lett.\  B {\bf 652} (2007) 34
  [arXiv:0705.0714 [hep-ph]];
  %%CITATION = PHLTA,B652,34;%%
P.~H.~Frampton and T.~W.~Kephart,
%  {\it Flavor Symmetry for Quarks and Leptons},
  JHEP {\bf 0709} (2007) 110
  [arXiv:0706.1186 [hep-ph]];
  %%CITATION = JHEPA,0709,110;%%
G.~J.~Ding,
%  {\it Fermion Mass Hierarchies and Flavor Mixing from $T'$ Symmetry},
  arXiv:0803.2278 [hep-ph].
  %%CITATION = ARXIV:0803.2278;%%



\bibitem{Lam:S4natural}
  C.~S.~Lam,
  %``The Unique Horizontal Symmetry of Leptons,''
  Phys.\ Rev.\   {\bf D\,78}, 073015 (2008),
  0809.1185.
  %%CITATION = PHRVA,D78,073015;%%



\bibitem{SU5xS4}
H.~Ishimori, Y.~Shimizu and M.~Tanimoto,
    %``S4 Flavor Symmetry of Quarks and Leptons in SU(5) GUT,''
    arXiv:0812.5031 [hep-ph].
    %CITATION = ARXIV:0812.5031;%%



\bibitem{bm}
F.~Bazzocchi and S.~Morisi,
  %``S4 as a natural flavor symmetry for lepton mixing,''
  arXiv:0811.0345 [hep-ph].
  %%CITATION = ARXIV:0811.0345;%%



\bibitem{S4Old}
S.~Pakvasa and H.~Sugawara,
    %``Mass Of The T Quark In SU(2) X U(1),''
    Phys.\ Lett.\  B {\bf 82} (1979) 105;
    %%CITATION = PHLTA,B82,105;%%
T.~Brown, N.~Deshpande, S.~Pakvasa and H.~Sugawara,
    %``CP Nonconservation And Rare Processes In S(4) Model Of Permutation
    %Symmetry,''
    Phys.\ Lett.\  B {\bf 141} (1984) 95;
    %%CITATION = PHLTA,B141,95;%%
T.~Brown, S.~Pakvasa, H.~Sugawara and Y.~Yamanaka,
    %``Neutrino Masses, Mixing And Oscillations In S(4) Model Of Permutation
    %Symmetry,''
    Phys.\ Rev.\  D {\bf 30} (1984) 255;
    %%CITATION = PHRVA,D30,255;%%
D.~G.~Lee and R.~N.~Mohapatra,
    %``An SO(10) x S(4) scenario for naturally degenerate neutrinos,''
    Phys.\ Lett.\  B {\bf 329} (1994) 463
    [arXiv:hep-ph/9403201];
    %%CITATION = PHLTA,B329,463;%%
E.~Ma,
    %``Neutrino mass matrix from S(4) symmetry,''
    Phys.\ Lett.\  B {\bf 632} (2006) 352
    [arXiv:hep-ph/0508231];
    %%CITATION = PHLTA,B632,352;%%
C.~Hagedorn, M.~Lindner and R.~N.~Mohapatra,
    %``S(4) flavor symmetry and fermion masses: Towards a grand unified theory  of
    %flavor,''
    JHEP {\bf 0606} (2006) 042
    [arXiv:hep-ph/0602244];
    %%CITATION = JHEPA,0606,042;%%
Y.~Cai and H.~B.~Yu,
    %``An SO(10) GUT Model with $S4$ Flavor Symmetry,''
    Phys.\ Rev.\  D {\bf 74} (2006) 115005
    [arXiv:hep-ph/0608022];
    %%CITATION = PHRVA,D74,115005;%%
F.~Caravaglios and S.~Morisi,
    %``Gauge boson families in grand unified theories of fermion masses: E_6^4   x
    %S_4,''
    Int.\ J.\ Mod.\ Phys.\  A {\bf 22} (2007) 2469
    [arXiv:hep-ph/0611078];
    %%CITATION = IMPAE,A22,2469;%%
H.~Zhang,
    %``Flavor S(4) x Z(2) symmetry and neutrino mixing,''
    Phys.\ Lett.\  B {\bf 655} (2007) 132
    [arXiv:hep-ph/0612214];
    %%CITATION = PHLTA,B655,132;%%
Y.~Koide,
    %``S_4 Flavor Symmetry Embedded into SU(3) and Lepton Masses and Mixing,''
    JHEP {\bf 0708} (2007) 086
    [arXiv:0705.2275 [hep-ph]].
    %%CITATION = JHEPA,0708,086;%%



\bibitem{FroggatNielsen}
C.~D.~Froggatt and H.~B.~Nielsen,
%  {\it Hierarchy Of Quark Masses, Cabibbo Angles And CP Violation},
  Nucl. Phys. B {\bf 147} (1979) 277.



\bibitem{HM}
  L.~Baudis {\it et al.},
  %``Limits on the Majorana neutrino mass in the 0.1 eV range,''
  Phys.\ Rev.\ Lett.\  {\bf 83}, 41 (1999)
  [arXiv:hep-ex/9902014].



\bibitem{gerda}
A.~A.~Smolnikov and f.~t.~G.~Collaboration,
  %``Status of the GERDA experiment aimed to search for neutrinoless double beta
  %decay of 76Ge,''
  arXiv:0812.4194 [nucl-ex].
  %%CITATION = ARXIV:0812.4194;%%



\bibitem{majorana}
{\it et al.}  [Majorana Collaboration],
  %``The Majorana Neutrinoless Double-Beta Decay Experiment,''
  arXiv:0811.2446 [nucl-ex].
  %%CITATION = ARXIV:0811.2446;%%



\bibitem{supernemo}
  H.~Ohsumi  [NEMO and SuperNEMO Collaborations],
  %``SuperNEMO project,''
  J.\ Phys.\ Conf.\ Ser.\  {\bf 120} (2008) 052054.
  %%CITATION = 00462,120,052054;%%




\bibitem{cuore}
  A.~Giuliani  [CUORE Collaboration],
  %``From Cuoricino to CUORE: Investigating the inverted hierarchy region of
  %neutrino mass,''
  J.\ Phys.\ Conf.\ Ser.\  {\bf 120} (2008) 052051.
  %%CITATION = 00462,120,052051;%%




\bibitem{exo}
  M.~Danilov {\it et al.},
  %``Detection of very small neutrino masses in double-beta decay using  laser
  %tagging,''
  Phys.\ Lett.\  B {\bf 480}, 12 (2000)
  [arXiv:hep-ex/0002003].



\bibitem{Kahler}
J.~R.~Espinosa and A.~Ibarra,
  %``Flavour symmetries and Kaehler operators,''
  JHEP {\bf 0408} (2004) 010
  [arXiv:hep-ph/0405095];
  %%CITATION = JHEPA,0408,010;%%
S.~F.~King, I.~N.~R.~Peddie, G.~G.~Ross, L.~Velasco-Sevilla and
O.~Vives,
% {\it Kaehler corrections and softly broken family symmetries},
  JHEP {\bf 0507} (2005) 049
  [arXiv:hep-ph/0407012];
  %%CITATION = JHEPA,0507,049;%%
S.~Antusch, S.~F.~King and M.~Malinsky,
  %``Third Family Corrections to Quark and Lepton Mixing in SUSY Models with
  %non-Abelian Family Symmetry,''
  JHEP {\bf 0805} (2008) 066
  [arXiv:0712.3759 [hep-ph]].
  %%CITATION = JHEPA,0805,066;%%



\bibitem{NoAlignment}
G.~Seidl,
  %``Unified model of fermion masses with Wilson line flavor symmetry
  %breaking,''
  arXiv:0811.3775 [hep-ph].
  %%CITATION = ARXIV:0811.3775;%%
W.~Grimus and L.~Lavoura,
  %``Tri-bimaximal lepton mixing from symmetry only,''
  arXiv:0811.4766 [hep-ph];
S.~Morisi,
  %``Tri-Bimaximal lepton mixing with A4 semidirect product Z2 x Z2 x Z2,''
  arXiv:0901.1080 [hep-ph].
  %%CITATION = ARXIV:0901.1080;%%


\end{thebibliography}
\end{document}